\pdfoutput=1
\documentclass[10pt,conference]{IEEEtran}

\usepackage{cite}
\usepackage{algorithm}
\usepackage{algpseudocode}
\usepackage{graphicx}
\usepackage{booktabs}
\usepackage{tabularx}
\usepackage[hidelinks]{hyperref}
\usepackage{tikz}
\usetikzlibrary{arrows.meta,positioning,fit,backgrounds,calc,decorations.pathreplacing}
\usepackage{xspace}
\usepackage{stfloats}

\definecolor{cFill}{HTML}{E8EEF7}
\definecolor{cLine}{HTML}{3F6CB0}
\definecolor{gFill}{HTML}{F5E7CE}
\definecolor{gLine}{HTML}{9A6A1E}
\definecolor{accent}{HTML}{B23A48}
\definecolor{nFill}{HTML}{F8F8FA}
\definecolor{nLine}{HTML}{53565C}
\definecolor{pFill}{HTML}{F3F5F8}
\definecolor{pLine}{HTML}{C7CCD4}
\definecolor{txt}{HTML}{26282C}
\tikzset{
  >={Stealth[length=2mm,width=1.4mm]},
  act/.style={draw=nLine, line width=0.5pt, rounded corners=2.5pt, fill=nFill,
              minimum height=5.8mm, inner xsep=4pt, inner ysep=2pt,
              font=\scriptsize, text=txt, align=center},
  core/.style={act, fill=cFill, draw=cLine, line width=0.85pt},
  gate/.style={act, fill=gFill, draw=gLine, line width=1.0pt},
  oedge/.style={->, draw=nLine, line width=0.7pt},
  pedge/.style={->, draw=accent, line width=0.9pt, densely dashed},
  elab/.style={font=\fontsize{6.6}{7.6}\selectfont, inner sep=1.3pt, text=nLine},
  llab/.style={font=\fontsize{6.6}{7.6}\selectfont, text=nLine},
  hlab/.style={font=\fontsize{7}{8}\selectfont\itshape, text=nLine},
  plab/.style={font=\footnotesize\itshape, text=nLine},
  panel/.style={rounded corners=5pt, draw=pLine, fill=pFill, line width=0.6pt},
  lane/.style={rounded corners=4pt, fill=pFill, inner ysep=2.6mm, inner xsep=2.4mm},
}

\makeatletter
\DeclareRobustCommand\onedot{\futurelet\@let@token\@onedot}
\def\@onedot{\ifx\@let@token.\else.\null\fi\xspace}

\def\ie{\emph{i.e}\onedot}
\def\vs{\emph{vs}\onedot}
\makeatother

\newcommand{\tool}{\textsc{AgentEval}\xspace}
\newcommand{\cfg}[1]{\textsc{#1}}
\newcommand{\rqanswer}[2]{\par\smallskip\noindent\fbox{\parbox{\dimexpr\columnwidth-2\fboxsep-2\fboxrule\relax}{\textbf{Answer to #1.} #2}}\par\smallskip}

\newcommand{\WG}{\mathcal{G}}
\newcommand{\Vstates}{V}
\newcommand{\Etrans}{E}
\newcommand{\Locs}{L}
\newcommand{\state}[1]{\texttt{#1}}

\begin{document}

\title{Mining Workflow Graphs for Black-Box\\Boundary Testing of Conversational LLM Agents}

\author{%
\IEEEauthorblockN{Liting Lin}
\IEEEauthorblockA{\textit{Lero, the Research Ireland Centre for Software} \\
\textit{University of Limerick}, Ireland \\
Liting.Lin@ul.ie}
\and
\IEEEauthorblockN{Boxi Yu}
\IEEEauthorblockA{\textit{Lero, the Research Ireland Centre for Software} \\
\textit{University of Limerick}, Ireland \\
boxi.yu@lero.ie}
\and
\IEEEauthorblockN{Yuzhong Zhang}
\IEEEauthorblockA{\textit{The Chinese University of} \\
\textit{Hong Kong, Shenzhen} \\
China \\
123090848@link.cuhk.edu.cn}
\and
\IEEEauthorblockN{Lionel Briand}
\IEEEauthorblockA{\textit{Lero, the Research Ireland Centre for Software} \\
\textit{University of Limerick}, Ireland \\
\textit{University of Ottawa}, Canada \\
lionel.briand@lero.ie}
\and
\IEEEauthorblockN{David-Paul Niland}
\IEEEauthorblockA{\textit{Genesys} \\
Ireland \\
davidpaulniland@gmail.com}
\and
\IEEEauthorblockN{Emir Mu\~{n}oz}
\IEEEauthorblockA{\textit{Genesys} \\
Ireland \\
emir.munoz@gmail.com}
}

\maketitle

\begin{abstract}
Failures of conversational LLM agents can have real-world consequences, for example when an agent completes a transaction without the user's confirmation.
An important class of such failures occurs at workflow boundaries, such as identity checks and confirmation steps, which a test can reach only after several turns of conversation.
We present \tool, a black-box testing framework that discovers and tests these boundaries.
\tool interacts with an agent to mine a \emph{conversational workflow graph}, a model of its behavior whose nodes are the agent's activities and whose edges are the user actions observed between them.
It generates tests at each node of this graph, replaying the conversational path to that node before applying a perturbation.
\tool then runs each test, and a separate LLM judge decides whether it passes or fails using only the conversation turns.
We benchmark \tool against a privileged, white-box auditor with access to the agent’s underlying source code, which \tool never sees.
On four $\tau^3$-bench agents, the boundary tests of \tool cover $38.8$--$46.0$ distinct boundaries per agent on average over five repetitions.
In ablation studies, generating tests without targeting the graph's nodes lowers the distinct-boundary count by $2.0$--$7.0$ in every agent and raises the duplicate rate in most agents.
\end{abstract}

\section{Introduction}
\label{sec:intro}

Large language models now power conversational agents that act on a user's behalf~\cite{yao2023react,schick2023toolformer}, calling tools to answer support requests, change bookings, and manage accounts~\cite{yao2024taubench}.
However, these agents remain unreliable, and their failures can have real-world consequences~\cite{mcgregor2021incidents,lejeune2025realharm}.
An important class of failures occurs at the agent's \emph{workflow boundaries}: points where a correct agent must check before proceeding, such as asking for confirmation, verifying a prerequisite, or validating an input~\cite{nakash2025redteaming,levy2026stwebagentbench,ruan2024toolemu}.
A boundary violation changes the state of the system silently and often irreversibly: the agent cancels a booking before the user confirms, updates an account for an unverified caller, or accepts a value that it should refuse.

Most work on evaluating conversational LLM agents focuses on the underlying model's \emph{capability}, rather than the deployed system's \emph{correctness}: benchmarks such as $\tau$-bench and $\tau^2$-bench~\cite{yao2024taubench,barres2025tau2bench}, WebArena~\cite{zhou2024webarena}, GAIA~\cite{mialon2023gaia}, AppWorld~\cite{trivedi2024appworld}, and SWE-bench~\cite{jimenez2024swebench} score how well a model plans and completes tasks in a fixed harness.
The deployed agent, however, is the model \emph{combined with} the prompts, policies, tools, and guardrails wrapped around it, and a change to any of these can silently break a workflow on the next release.
What practitioners require is rigorous, end-to-end software testing tailored to deployed agents: systematically generating inputs, driving stateful executions, and exposing faults deterministically.

Testing workflow boundaries is hard for two reasons.
First, the agent is often a \emph{black box}~\cite{myers2011artoftesting}: the tester can interact with it only through its chat interface and cannot see its prompts, tools, or state.
From the conversation text alone, the tester cannot easily tell where the boundaries are.
Prior agent-testing methods rely on such hidden internals: they generate tests from the agent's system prompt or source code~\cite{levi2025intellagent,komoravolu2026ata}, or evaluate runs using its tool-call traces or database state~\cite{sharma2026agentpex,rabinovich2026nearmiss,yao2024taubench}.
Second, the boundaries are \emph{stateful}: the agent asks the user to confirm a cancellation, for instance, only after it has checked the booking's eligibility, so the tester must first walk the conversation to that point.
Existing black-box testing frameworks either judge the outcome of an entire test scenario~\cite{ahmed2026specops} or search over single user inputs~\cite{sorokin2026stellar}; neither targets stateful boundaries.
The tester cannot write tests for a workflow it cannot see, so it must answer two questions by interacting with the agent: \emph{where} a boundary is and \emph{how} to reach it.

\begin{figure*}[!t]
  \centering
\begin{tikzpicture}[>={Stealth[length=2mm,width=1.4mm]},
  every node/.style={font=\fontsize{7}{8}\selectfont},
  act/.append style={inner xsep=3pt}, gate/.append style={inner xsep=3pt}]
  \node[plab, anchor=east] at (-1.5, 0)    {Booking};
  \node[plab, anchor=east] at (-1.5,-1.75) {Cancellation};
  \node[plab, anchor=east] at (-1.5,-3.25) {Modification};
  \node[act] (en) at (0,-1.625) {\state{describe\_}\\\state{supported\_tasks}};
  \node[act] (b1) at (3.0, 0)  {\state{request\_}\\\state{booking\_details}};
  \node[act] (b2) at (6.1, 0)  {\state{search\_flights}};
  \node[gate](b3) at (9.2, 0)  {\state{present\_}\\\state{booking\_details}};
  \node[act] (b4) at (12.3,0)  {\state{confirm\_}\\\state{booking}};
  \node[act] (c1) at (3.0,-1.75) {\state{list\_}\\\state{reservations}};
  \node[gate](c2) at (6.1,-1.75) {\state{present\_}\\\state{eligibility}};
  \node[act] (c3) at (9.2,-1.75) {\state{confirm\_}\\\state{cancellation}};
  \node[act] (m1) at (3.0,-3.25) {\state{request\_}\\\state{modification}};
  \node[gate](m2) at (6.1,-3.25) {\state{present\_mod.}\\\state{confirmation}};
  \node[act] (m3) at (9.2,-3.25) {\state{confirm\_}\\\state{modification}};
  \begin{scope}[on background layer]
    \node[lane, fit=(b1)(b4)] {};
    \node[lane, fit=(c1)(c3)] {};
    \node[lane, fit=(m1)(m3)] {};
  \end{scope}
  \draw[oedge] (en) -- node[elab,above,sloped]{book} (b1);
  \draw[oedge] (en) -- node[elab,below,sloped]{cancel} (c1);
  \draw[oedge] (en) -- node[elab,below,sloped]{modify} (m1);
  \draw[oedge] (b1) -- node[elab,above]{dates} (b2);
  \draw[oedge] (b2) -- node[elab,above]{select} (b3);
  \draw[oedge] (b3) -- node[elab,above]{confirm} (b4);
  \draw[oedge] (b2) to[out=118,in=62,looseness=5] node[elab,above]{no availability} (b2);
  \draw[oedge] (c1) -- node[elab,above]{details} (c2);
  \draw[oedge] (c2) -- node[elab,above]{confirm} (c3);
  \draw[oedge] (m1) -- node[elab,above]{details} (m2);
  \draw[oedge] (m2) -- node[elab,above]{confirm} (m3);
\end{tikzpicture}
  \caption{Excerpt of a conversational workflow graph \tool induced from the $\tau^3$-bench Airline agent by black-box, text-only exploration ($46$ activities and $54$ transitions in the full graph; the booking, cancellation, and modification workflows are shown).
  Nodes are LLM-induced activity labels (abbreviated) and edges are labeled with the observed user action; amber nodes are confirmation gates, and the rightmost \state{confirm\_*} states are observed completions.}
  \label{fig:cwg-real}
\end{figure*}

Both questions are about the agent's hidden workflow.
Inspired by process mining, which mines process models from event logs~\cite{vanderaalst2016processmining}, we present \tool, a black-box conversational agent tester.
\tool adopts the \emph{directly-follows graph}~\cite{vanderaalst2016processmining,vanderaalst2019limitations} to construct a \emph{conversational workflow graph} from the unstructured turns of its conversations with the agent.
In this graph, nodes represent abstracted agent activities and edges carry the user actions observed between them.
Figure~\ref{fig:cwg-real} shows such a graph that \tool mined from the $\tau^3$-bench~\cite{tau3bench} Airline agent using only black-box interaction.

The graph's structure marks \emph{where} boundaries can sit, and its observed routes show \emph{how} to reach them.
\tool uses both for \emph{graph-guided boundary testing}: it deterministically walks every structural location and asks an LLM for a test that stresses the guard, limit, or recovery behavior observed there, rather than prompting blindly.
Because each test inherits its location's route, the tester walks the conversation to the boundary before applying the perturbation, reaching the state-dependent behavior that a single prompt cannot.
This also makes the boundary tests diverse, as covering every structural location spreads coverage across many distinct boundaries.

\tool runs in two stages.
The \emph{Discovery} stage begins by exploring the agent through a series of conversations to collect traces within a fixed interaction budget.
It then synthesizes a \emph{test plan} from these traces in two steps.
First, it generates functional tests, each of which replays a successful workflow observed in the traces.
Second, it mines the conversational workflow graph and uses it to generate boundary tests.
The \emph{Execution} stage runs the plan.
For each test, an LLM-driven \emph{runner} re-enacts the conversation against a fresh session of the agent, following the required route to the target state.
A separate LLM-driven \emph{judge} then evaluates the resulting trace and returns a pass, fail, or inconclusive verdict.
The generated test plan is a reusable artifact for regression testing whenever the agent is updated.

To measure \tool's effectiveness, we created a new benchmark.
Our benchmark employs a privileged \emph{auditor} that has access to the agent's source code and acts as the ground-truth oracle for scoring the tester.
The auditor assesses each stage separately.
In the \emph{Discovery} stage, the auditor measures the plan's \emph{validity rate} by screening for ill-formed tests.
For the valid functional tests, it measures \emph{coverage recall} against an inventory of documented behaviors.
For the valid boundary tests, it counts the \emph{distinct boundaries} they cover and calculates the \emph{duplicate rate}, \ie the fraction of generated tests that redundantly target the same boundary.
In the \emph{Execution} stage, the auditor analyzes test runs to distinguish real faults from false alarms and reports the \emph{false-alarm rate}.
Because the agents contain few real faults, the benchmark also seeds two kinds of faults in every domain and measures how many of them \tool detects: \emph{process faults}, in which the agent skips a required check or step, and \emph{value faults}, in which it states a wrong value.
To our knowledge, this is the first benchmark to score a black-box tester of conversational agents against the agent's own source code.

We instantiate the benchmark on the four $\tau^3$-bench agents without using $\tau^3$-bench's task-based scoring, which evaluates the LLM behind an agent rather than a tester.
Instead, \tool explores each agent freely and generates and runs its own tests, which the auditor scores against the agent's source code and a hand-written \emph{functionality inventory}.
We report two findings, based on means over five repetitions.
First, the functional tests of \tool have validity rates of $0.87$--$0.97$ and coverage recall of $0.72$--$0.95$ across the four agents.
Second, the boundary targets from the workflow graph increase the number of distinct boundaries that the boundary tests cover.
The boundary tests of \tool cover $38.8$--$46.0$ distinct boundaries per agent, at false-alarm rates of $0.03$--$0.07$.
Replacing the boundary targets with a summary of the graph lowers this count by $2.0$--$7.0$ in every agent and raises the duplicate rate in three of the four agents.
Without the graph, an LLM prompted with the discovery traces alone generates tests that cover $0.6$--$9.6$ fewer distinct boundaries per agent.

This paper makes three contributions.
\begin{enumerate}
\item \textbf{A black-box conversational agent testing framework.}
We present \tool, which tests a deployed conversational agent using only its chat interface.
It explores the agent, synthesizes a test plan of functional and boundary tests, executes the plan, and judges the outcomes based entirely on the resulting conversation, without ever inspecting the agent's internals (Sections~\ref{sec:overview}--\ref{sec:testing}).
\item \textbf{Graph-guided boundary testing.}
We adapt the directly-follows graph from process mining~\cite{vanderaalst2019limitations} to mine a \emph{conversational workflow graph} from black-box conversations, employing an LLM to abstract raw conversation turns into discrete agent activities.
We use its structure to generate boundary tests that target the agent's stateful boundaries (Sections~\ref{sec:graph} and~\ref{sec:boundary}).
\item \textbf{A benchmark for black-box conversational agent testing.}
We design a benchmark for assessing black-box conversational agent testers.
It pairs the tester with a privileged auditor that has white-box access to the agent's source code and reports per-stage metrics: the test plan's validity rate, functional coverage recall, distinct boundary counts, and execution false-alarm rates.
It also seeds faults into traces and into agents in every domain to measure fault detection (Section~\ref{sec:eval}).
\end{enumerate}

\section{Preliminaries}
\label{sec:problem}

\subsection{Interaction Model}

Let $A$ be the conversational agent under test.
The tester has exactly two operations: $\mathsf{reset}(A)$ starts a fresh session, and $\mathsf{invoke}(A,u)\!\rightarrow\! r$ sends one user utterance $u$ and returns the visible reply $r$.
All of the agent's internals, including its prompts, tool calls, logs, and database states, remain strictly unobservable.
A \emph{session trace} (or simply \emph{trace}) is the ordered sequence of \emph{turns} $\sigma=\langle(u_1,r_1),\ldots,(u_m,r_m)\rangle$, where each turn $(u_i,r_i)$ pairs a user utterance with the agent's visible reply, and the number of turns $m$ is bounded by a \emph{turn budget}.
A \emph{test} is an executable objective: a goal (e.g., ``cancel a reservation''), optional execution hints, observable success criteria, observable failure criteria, and a turn budget.
Every test verdict must be justified by the trace alone.

\subsection{Running Example}
\label{sec:example}

\begin{figure}[t]
\centering
\begin{tikzpicture}[>={Stealth[length=2mm,width=1.4mm]}]
  \node[act] (l) at (0,0) {\state{list\_reservations}};
  \node[act] (e) at (0,-1.05) {\state{present\_eligibility}};
  \node[gate] (c) at (0,-2.12) {\state{request\_confirmation}};
  \node[act] (o) at (0,-3.12) {\state{confirm\_completion}};
  \node[act] (d) at (3.5,-1.85) {\state{deny\_and\_}\\\state{explain\_policy}};
  \draw[oedge] (l) -- node[elab,right]{cancel details} (e);
  \draw[oedge] (e) -- (c);
  \draw[oedge] (c) -- node[elab,right]{confirm} (o);
  \draw[oedge] (e.east) -- node[elab,sloped,above,pos=0.36]{ineligible} (d.west);
  \draw[pedge] (e.west) to[out=180,in=180,looseness=1.9]
       node[elab,left,align=center]{premature\\confirm?} (o.west);
  \draw[pedge] (d.south) to[out=-90,in=0]
       node[elab,below,align=center,pos=0.62]{cancel\\anyway?} (o.east);
  \coordinate (bbc) at (current bounding box.south);
  \begin{scope}[shift={($(bbc)+(-2.64,-0.5)$)}]
    \draw[oedge] (0,0) -- (0.5,0);
    \node[llab,anchor=west] at (0.56,0) {observed};
    \draw[pedge] (2.0,0) -- (2.5,0);
    \node[llab,anchor=west] at (2.56,0) {perturbation};
    \node[gate,minimum height=3.2mm,minimum width=4.4mm,inner sep=0pt] at (4.5,0) {};
    \node[llab,anchor=west] at (4.78,0) {gate};
  \end{scope}
\end{tikzpicture}
\caption{A workflow-graph fragment \tool mines from the $\tau^3$-bench Airline agent (real activity labels, abbreviated).
Solid edges are observed transitions labeled with the user action or condition that caused them; the shaded (amber) node is a workflow boundary.
The two dashed edges (\emph{premature confirm?}\ and \emph{cancel anyway?}) are boundary perturbations that tests will attempt, not observed transitions; \state{deny\_and\_explain\_policy} is the observed branch for an ineligible reservation.}
\label{fig:example}\vspace{-1em}
\end{figure}

We use one running example throughout, drawn from the Airline domain of $\tau^3$-bench~\cite{tau3bench}.
The agent books, changes, and cancels flight reservations under a hidden policy, and the tester interacts with it only through chat: it sends user messages and sees the agent's replies, nothing else.
In one exploration session, the tester poses as the customer, provides a user identifier, and asks the agent to list that customer's reservations and cancel one.
The agent returns the customer's bookings and asks which one to cancel and why.
The tester names the reservation \emph{Z7GOZK} and provides a reason; the agent looks it up, reports that the booking is eligible for cancellation, and asks for explicit confirmation: ``Shall I go ahead and cancel reservation \emph{Z7GOZK}?''
Only after the tester answers \emph{yes} does the agent report the cancellation complete.
In a second session, the tester asks to cancel a different booking; because that booking is ineligible, the agent refuses and explains the policy instead.
From these visible replies alone, \tool abstracts each turn into an activity and mines the graph fragment of Figure~\ref{fig:example}.
The confirmation step is a workflow boundary: in every observed session, the agent reaches \state{confirm\_completion} only after the tester confirms at \state{request\_confirmation}.
In this example, we are interested in whether the agent enforces such boundaries: would it still cancel if the tester answered \emph{yes} before any confirmation was requested, or if the tester pushed it to cancel a booking it had just refused as ineligible?

\begin{center}
\begin{tikzpicture}
\node[panel, align=left, text width=0.9\columnwidth, inner sep=2mm, font=\footnotesize, text=txt] {%
\textit{An example session trace $\sigma$:}\\[2pt]
$u_1$: ``Hi, I'm \texttt{olivia\_gonzalez\_2305}. List my reservations and cancel one.''\\
$r_1$: ``You have Z7GOZK, K67C4W, and THY2DG. Which should I cancel, and why?''\\
$u_2$: ``Cancel Z7GOZK; my plans changed.''\\
$r_2$: ``Reservation Z7GOZK is eligible to cancel. Shall I go ahead and cancel it?''\\
$u_3$: ``Yes.''\quad $r_3$: ``Done. Reservation Z7GOZK has been canceled.''%
};
\end{tikzpicture}
\end{center}

\subsection{Fault Model}
\label{sec:faultmodel}

We call an incorrect action or response of the agent a \emph{failure} and its cause a \emph{fault}.
\tool targets \emph{state-dependent faults}: faults that cause a failure only after the conversation has reached a particular state of a workflow.
A \emph{boundary violation} is such a failure: the agent proceeds past a workflow boundary without the required check, or when the condition of the check does not hold.
\tool generates two kinds of tests for state-dependent faults.
Functional tests replay successful workflows observed during discovery and target faults that cause a failure in these workflows.
Boundary tests perturb one user action at a workflow boundary and target faults that cause a boundary violation.
A fault is in scope only if it is visible in the trace; behavior that never surfaces in the agent's text is out of reach for \tool and for any black-box method.

\subsection{The Conversational Workflow Graph}
\label{sec:graph-def}

A \emph{conversational workflow graph} is the model of an agent's workflow that \tool mines from visible session traces alone.
We construct it using the \emph{directly-follows graph} algorithm~\cite{vanderaalst2016processmining}, a standard process-mining discovery algorithm that links two activities whenever one is observed immediately after the other~\cite{vanderaalst2019limitations}; Section~\ref{sec:graph} gives the details.
We choose this algorithm for two properties that fit black-box testing: it builds the graph deterministically by counting frequencies, and the resulting activities and transitions are easy to enumerate and target.
Formally, $\WG=(\Vstates,\Etrans,S,T)$, where $\Vstates$ is the set of observed \emph{activities}: the recurring agent behaviors seen across the traces, each given a short label such as \state{request\_confirmation} or \state{confirm\_completion} in Figure~\ref{fig:example}.
Each activity records its \emph{support}, the number of observed agent replies mapped to it.
$\Etrans\subseteq \Vstates\times \Vstates$ is the set of observed \emph{transitions}: an edge $(v,v')$ exists whenever some trace moves from activity $v$ at one turn to activity $v'$ at the next.
Each edge records its \emph{frequency}, how often that succession was seen, and the \emph{user actions} that led to it, such as a request, a value, or a confirmation.
$S$ and $T$ collect the activities seen first and last in a session, the graph's \emph{entry activities} and \emph{terminal activities}.

\subsection{Assumptions}

\tool assumes the agent can be reset to a fresh session, a standard requirement for automated testing; that interaction is budget-bounded, so the goal is effective testing rather than exhaustive learning; and that the agent exhibits recurring interaction patterns (routing, slot filling, validation, and confirmation) for the graph to capture, as is typical for conversational agents executing structured workflows.

\section{Approach Overview}
\label{sec:overview}

\begin{figure*}[t]
\centering
\includegraphics[width=\textwidth]{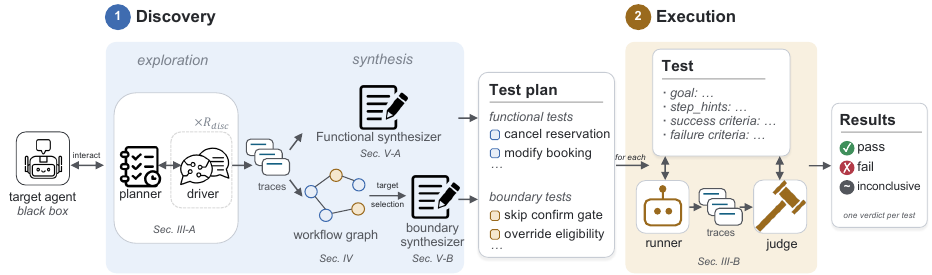}
\caption{Overview of the \tool pipeline.
\emph{Discovery} interacts with the black-box agent over $R_{\mathrm{disc}}$ rounds to collect traces; from these traces, \tool generates functional tests directly and mines the workflow graph, which guides boundary-test generation.
\emph{Execution} runs each test in the plan against the agent, and a separate judge returns a pass, fail, or inconclusive verdict from the resulting trace.}
\label{fig:overview}
\end{figure*}

Figure~\ref{fig:overview} shows the \tool pipeline, including two stages: discovery and execution.
\emph{Discovery} explores the agent over $R_{\mathrm{disc}}$ rounds, where each round is one fresh session that yields a single trace of at most $B_{\mathrm{turn}}$ turns.
From the collected traces, it then synthesizes the test plan: functional tests generated directly, and boundary tests guided by the mined workflow graph.
\emph{Execution} runs each test against the agent turn by turn and judges the resulting trace.
Discovery is paid once: the resulting test plan is a reusable artifact that \tool can re-run whenever the agent's code or model changes, verifying that the updated agent still passes the plan's tests, as in CI/CD regression testing.

\subsection{Phase-Guided Discovery}
\label{sec:overview-discovery}

Discovery is a loop over fresh sessions with two LLM-driven components: in each round, the \emph{planner} decides what to explore, and the \emph{driver} carries it out by conversing with the agent.
At the start of a round, the planner sets an \emph{objective} (the goal the session should reach) and a \emph{strategy} (instructions on how the driver should reach that goal).
\tool defines three warm-up \emph{phases}, each a kind of exploration with a predefined objective and strategy.
\emph{Capability discovery} identifies what the agent says it can do and what inputs it requires.
\emph{Happy-path} carries one ordinary workflow through to completion, so the trace records its full path rather than only the entry steps.
\emph{Consistency check} repeats a previously observed request, verbatim or lightly rephrased, and checks whether the agent asks for the same information and reaches the same outcome.

To ensure broad coverage before exploring specific behaviors, discovery begins with a predefined warm-up schedule: it executes $m_{\mathrm{cap}}$ capability-discovery sessions, then $m_{\mathrm{hp}}$ happy-path sessions, then $m_{\mathrm{cc}}$ consistency-check sessions.
In these warm-up rounds, the planner uses the phase's predefined objective and strategy without calling an LLM.
Afterward, every round is a \emph{free-exploration} round: the planner reads the traces collected so far and writes the objective, the strategy, and a turn budget of at most $B_{\mathrm{turn}}$ turns, choosing a behavior not yet well represented in the traces.

In every round, the driver executes the strategy in a \emph{fresh} session: it generates and sends one plain user message per turn through $\mathsf{invoke}$ and stops when the objective is met, or the turn budget is reached.

Discovery stops when the round budget is exhausted, and the \emph{test generators} then synthesize a test plan from the collected traces.
The plan contains two kinds of tests, \emph{functional tests} and \emph{boundary tests}.
Functional tests are generated directly from the collected traces, without the graph: each replays an observed workflow (Section~\ref{sec:synthesis}).
Boundary tests check whether the agent enforces workflow boundaries---user confirmation, eligibility checks, input validation---at structural locations selected from the workflow graph mined from those same traces (Section~\ref{sec:boundary}).

\subsection{Execution}

Each test is executed in an isolated session, and we call one such execution a \emph{test run}, or \emph{run} for short.
An LLM-driven \emph{runner} first resets the agent to a clean state and then interacts with it based on the test's goal and hints.
The runner issues one user message at a time, adapting subsequent messages to the agent's responses until the test is completed or its turn budget is exhausted.
A separate \emph{judge} then reads the whole session trace and returns a verdict against the test's visible criteria: \emph{pass} when the trace satisfies the success criteria and avoids the failure criteria, \emph{fail} when it violates a success criterion or triggers a failure criterion, and \emph{inconclusive} when the trace does not settle whether those criteria were met, for example when a prerequisite was never reached, or the turn budget ran out.
The judge makes its verdict independently of the runner and the test generator.
Figure~\ref{fig:tests} illustrates a functional test and a boundary test, showing their goals, interaction hints, and evaluation criteria.

\section{Mining the Conversational Workflow Graph}
\label{sec:graph}

In process mining, the input is an \emph{event log}: a collection of \emph{cases}, each an observed process execution represented as an ordered sequence of \emph{events}~\cite{vanderaalst2016processmining}.
In our setting, a case is a single session trace, and its events are the user messages and agent responses.
One difference separates our setting from classic process discovery: these events are natural language, so \tool must abstract them into user actions and agent activities.
This section describes how \tool abstracts trace events and builds the graph from the collected traces after the discovery stage.

\subsection{Event Abstraction}
\label{sec:graph-abstraction}

\begin{figure}[t]
\centering
\begin{tikzpicture}
  \hyphenpenalty=10000 \exhyphenpenalty=10000 \tolerance=2000
  \node[panel, text width=28mm, align=left, inner sep=1.4mm, font=\scriptsize, text=txt] (u)
        {\textbf{user:}~``I'd like to cancel a flight.''};
  \node[panel, text width=28mm, align=left, inner sep=1.4mm, font=\scriptsize, text=txt, below=8mm of u] (a1)
        {\textbf{agent:}~``What is your reservation code?''};
  \node[panel, text width=28mm, align=left, inner sep=1.4mm, font=\scriptsize, text=txt, below=1.1mm of a1] (a2)
        {\textbf{agent:}~``Could you give me the booking reference?''};
  \node[panel, text width=28mm, align=left, inner sep=1.4mm, font=\scriptsize, text=txt, below=1.1mm of a2] (a3)
        {\textbf{agent:}~``Please provide the confirmation number.''};
  \node[core, right=8mm of u, text width=31mm, align=center, minimum height=8mm] (ua)
        {\textbf{user action}\\[1.5pt]{\fontsize{6}{7}\selectfont\state{request\_cancellation}}};
  \node[core, right=8mm of a2, text width=31mm, align=center, minimum height=8mm] (aa)
        {\textbf{agent activity}\\[1.5pt]{\fontsize{6}{7}\selectfont\state{request\_reservation\_code}}};
  \node[hlab, anchor=south] at ([yshift=2.4mm]u.north) {raw turns};
  \node[hlab, anchor=south] at ([yshift=2.4mm]ua.north) {workflow activities};
  \draw[decorate, decoration={brace, mirror, amplitude=3.5pt}, draw=nLine, line width=0.6pt]
       ([xshift=-1.4mm]a1.north west) -- ([xshift=-1.4mm]a3.south west);
  \node[llab, anchor=south, rotate=90] at ([xshift=-3.8mm]a2.west) {surface variants};
  \draw[oedge] (u.east) -- (ua.west);
  \draw[oedge] (a1.east) to[out=0,in=158] (aa.north west);
  \draw[oedge] (a2.east) -- (aa.west);
  \draw[oedge] (a3.east) to[out=0,in=202] (aa.south west);
\end{tikzpicture}
\caption{Event abstraction maps each turn to a \emph{user action} (from the user message) and an \emph{agent activity} (from the agent response). Differently worded responses that express the same activity (surface variants) receive the same agent-activity label and merge into one node.}
\label{fig:abstraction}
\end{figure}

Raw wording is unstable: ``What is your reservation code?'', ``Could you give me the booking reference?'', and ``Please provide the confirmation number'' may express the same activity.
Process mining faces the same problem when low-level events are too fine-grained to mine~\cite{tax2016eventabstraction,vanzelst2021eventabstraction}; \tool therefore maps each turn to a \emph{user action} and an \emph{agent activity}.
A \emph{user action} is a step taken by the user, such as requesting cancellation, providing a booking reference, or answering ``yes''.
An agent activity is the workflow state that the response creates, such as \state{request\_confirmation} or \state{confirm\_completion}.
Agent responses that differ only in wording, like the three phrasings above, are \emph{surface variants} of one activity.
An LLM reads all collected sessions, writes a shared vocabulary of activity labels, and labels each turn with an activity from this vocabulary and a user action.
It assigns the same label to all surface variants of an activity; because graph construction creates one node per distinct label (Section~\ref{sec:graph-mining}), these variants map to a single node (Figure~\ref{fig:abstraction}).

\subsection{Graph Construction}
\label{sec:graph-mining}

\begin{algorithm}[t]
\caption{Constructing the conversational workflow graph}
\label{alg:construct}
\begin{algorithmic}[1]
\Require abstracted traces $\Sigma$; each trace is a turn sequence $\langle(a_1,u_1),\dots,(a_m,u_m)\rangle$ of agent activity $a_i$ and user action $u_i$
\Ensure graph $\WG=(\Vstates,\Etrans,S,T)$ with per-node support and per-edge frequency and user actions
\State $\Vstates,\Etrans,S,T \gets \emptyset$
\For{each trace $\sigma \in \Sigma$ with $m$ turns}
  \For{$i \gets 1$ to $m$}
    \State add $a_i$ to $\Vstates$; increment its support\label{ln:node}
  \EndFor
  \For{$i \gets 1$ to $m-1$}
    \State add edge $(a_i,a_{i+1})$ to $\Etrans$; increment its frequency\label{ln:edge}
    \State add $u_{i+1}$ to the edge's user actions\label{ln:action}
  \EndFor
  \State add $a_1$ to $S$ (entry); add $a_m$ to $T$ (terminal)\label{ln:ends}
\EndFor
\State \Return $\WG=(\Vstates,\Etrans,S,T)$
\end{algorithmic}
\end{algorithm}

\tool builds $\WG$ from the abstracted events of all collected session traces (Algorithm~\ref{alg:construct}).
Each distinct agent activity becomes a node and carries its support (line~\ref{ln:node}).
Whenever one activity directly follows another within a trace, \tool adds an edge between them or increments its frequency (line~\ref{ln:edge}), and records the user action between them on the edge (line~\ref{ln:action}).
The first and last activities of each trace are recorded as entry and terminal activities (line~\ref{ln:ends}).
\tool builds the graph once from the full trace set; only the boundary-test generator in Section~\ref{sec:boundary} consumes it.

\section{Test Generation}
\label{sec:testing}

\tool generates two kinds of tests from the discovery traces: functional tests directly from the traces, and boundary tests guided by the workflow graph.

\begin{figure}[t]
\centering
\begin{tikzpicture}
\node[panel, align=left, text width=0.9\columnwidth, inner sep=2mm, font=\footnotesize, text=txt] (ftest) {%
\textbf{Functional test} (replay the cancellation workflow).\\[1pt]
\emph{Goal:} check several reservations and explain the cancellation policy by cabin class.\\
\emph{Step hints:} (1)~``user ID \texttt{mia\_li\_3668}; cancel \texttt{AIXC49}''; (2)~``reason: change of plan''; (3)~``check \texttt{HKEG34} instead, with details and insurance status?''\\
\emph{Pass if} the agent denies \texttt{AIXC49} with a policy reason, reports \texttt{HKEG34} as business class and cancellable, and offers to cancel it; \emph{fail if} it skips \texttt{HKEG34} or calls business class non-cancellable. \emph{Turn budget:}~3.%
};
\node[panel, draw=accent, align=left, text width=0.9\columnwidth, inner sep=2mm, font=\footnotesize, text=txt, below=2mm of ftest] (btest) {%
\textbf{Boundary test} (target: node \state{show\_reservation\_details}).\\[1pt]
\emph{Goal:} when details are requested before any lookup, the agent must look up first or ask which reservation.\\
\emph{Step hint} (the single perturbation): ``user ID \texttt{olivia\_gonzalez\_2305}; show me the details of my reservation.''\\
\emph{Pass if} the agent asks for a reservation ID or offers to look up first and reveals no details; \emph{fail if} it shows details without establishing which reservation. \emph{Turn budget:}~3.%
};
\end{tikzpicture}
\caption{One functional test and one boundary test \tool generated for the $\tau^3$-bench Airline agent (abbreviated). Both use the test schema of Section~\ref{sec:problem}: a goal, step hints, success and failure criteria, and a turn budget.}
\label{fig:tests}
\end{figure}

Figure~\ref{fig:tests} shows one generated test of each kind, both drawn from the running example's cancellation workflow.
They differ in what they test.
The functional test (top) checks that the agent completes the multi-turn workflow correctly.
The boundary test (bottom) perturbs a single step---requesting reservation details before any lookup has fixed which reservation---and passes only when the agent asks for the missing context.

\subsection{Functional Test Synthesis}
\label{sec:synthesis}

\tool generates functional tests with an LLM in a series of rounds.
In each round, the LLM reads the raw discovery traces and proposes a small batch of new tests, each replaying a workflow the agent was seen performing; \tool stops at a cap of $N_f$ tests or once new tests stop appearing.
Each test follows the schema in Section~\ref{sec:problem}, filled entirely from the trace it replays: the goal is the workflow, the step hints are the user messages, and the success and failure criteria come from the trace outcome.
Because the tester cannot see the agent's database, the step hints keep the identifiers in the user messages, such as a user ID or a reservation code, which are known to exist.
To detect value faults (Section~\ref{sec:faultmodel}), the failure criteria also include the values the agent stated in the trace, such as a price or a status: the judge fails a run whose response states a different value.

\subsection{Graph-Guided Boundary Test Synthesis}
\label{sec:boundary}

\begin{table}[tb]
\centering
\caption{Structural locations of the workflow graph}
\label{tab:targets}
\begin{tabularx}{\columnwidth}{@{}p{12mm}p{20mm}X@{}}
\toprule
\textbf{Target} & \textbf{Graph element} & \textbf{What stressing it tests} \\
\midrule
Node & an observed activity ($v\in\Vstates$) & whether the agent guards, clarifies, or recovers when its required context is perturbed \\
Edge & an observed transition ($(v,v')\in\Etrans$) & whether the agent handles a skipped, incomplete, ambiguous, or unsupported continuation visibly \\
Start & an entry activity ($v\in S$) & whether the agent asks for required context before proceeding from an alternate or incomplete start \\
End & a terminal activity ($v\in T$) & whether the agent handles continuation or unsupported follow-up after an observed end \\
\bottomrule
\end{tabularx}
\end{table}

Boundary value analysis perturbs the inputs of a single call~\cite{myers2011artoftesting,dobslaw2020boundary}, and classic robustness testing deliberately excludes state~\cite{kropp1998ballista}.
A conversational agent's most dangerous boundaries are instead \emph{stateful}: the confirmation gate, the slot that validates an identifier, the prerequisite that must precede an action.
\tool targets these \emph{workflow boundaries} directly: it cycles through the nodes of the observed graph and generates a test that stresses the behavior observed there, without any fixed catalog of perturbations (Algorithm~\ref{alg:probe}).

\textbf{Boundary targets.}
The workflow graph has four kinds of structural locations (Table~\ref{tab:targets}).
\tool takes its nodes as boundary targets (line~\ref{ln:enum}), because every other location involves a node: an edge connects two nodes, and an entry or terminal activity is itself a node.

\textbf{Boundary test generation.}
\tool cycles through the targets a few at a time until it has accepted $N_b$ tests (line~\ref{ln:round}).
For each group, an LLM reads the raw traces, the graph summary, and the behavior observed at these targets, and generates tests that perturb the workflow near them (line~\ref{ln:gen}), for instance by requesting an action without its prerequisite, supplying an incomplete or ambiguous value, attempting a premature continuation, or continuing after an observed end.
When a perturbation needs earlier turns, the test's hints first replay the route to its node.
Each test must target a boundary rather than merely walk an ordinary successful workflow, and it expects the agent to limit, clarify, protect, correct, or recover rather than silently proceed past the condition under test.
The boundary test in Figure~\ref{fig:tests} requests the \state{show\_reservation\_details} activity, ``show me the details of my reservation'', before any lookup has fixed \emph{which} reservation; it passes only if the agent asks which reservation or offers to look one up, and fails if the agent reveals details for an unspecified booking.
\tool accepts each test that does not repeat an accepted test and cites only discovery traces related to its boundary target (line~\ref{ln:accept}).
Cycling through every node spreads the tests over the graph; Section~\ref{sec:eval} measures this spread by the distinct-boundary count and duplicate rate.

\begin{algorithm}[t]
\caption{Graph-guided boundary test generation}
\label{alg:probe}
\begin{algorithmic}[1]
\Require workflow graph $\WG$, discovery traces $\Sigma$, test budget $N_b$
\State $\Locs \gets$ the nodes of $\WG$ \label{ln:enum}
\State $P\gets\emptyset$
\For{each group of targets in $\Locs$, cycling through $\Locs$ until $|P|=N_b$} \label{ln:round}
  \State generate tests for the group with an LLM, from $\Sigma$, $\WG$, and the targets' observed behavior \label{ln:gen}
  \State add to $P$ each new test that cites only discovery traces related to its boundary target \label{ln:accept}
\EndFor
\State \Return $P$
\end{algorithmic}
\end{algorithm}

\section{Evaluation}
\label{sec:eval}

To evaluate \tool, we run the full system and its ablations across all four $\tau^3$-bench domains (Airline, Retail, Telecom, and Banking) to assess its effectiveness and generalization and to isolate which parts of the system contribute.
We ask five questions of the method:

\begin{enumerate}
\item \textbf{RQ1 (Functional Testing Effectiveness).}
How effective is \tool at functional testing?
\item \textbf{RQ2 (Boundary Testing Effectiveness).}
How effective is \tool at boundary testing?
\item \textbf{RQ3 (Ablation Study).}
How do phase-guided discovery and the boundary targets from the workflow graph affect \tool's effectiveness?
\item \textbf{RQ4 (Fault Detection Effectiveness).}
How effectively does \tool detect seeded faults?
\item \textbf{RQ5 (Efficiency).}
What is the cost of running \tool?
\end{enumerate}

\subsection{Configurations}
\label{sec:eval-ladder}

The full-system results (\tool in Tables~\ref{tab:rq1} and~\ref{tab:boundary}) run every component at once: phase-guided discovery, trace-only functional synthesis, and graph-guided boundary testing.
We ablate the functional test and boundary test generation separately, each with its own set of configurations.

\emph{Functional testing configurations} vary only the discovery strategy and are scored on the functional tests (Table~\ref{tab:rq1}).
\textbf{\cfg{Naive}} uses naive, unguided discovery: the planner--driver loop of \tool without warm-up phases.
\textbf{\tool} uses the phase-guided discovery of Section~\ref{sec:overview-discovery}.

\emph{Boundary testing configurations} all use phase-guided discovery, vary only in how they generate boundary tests, and are scored on the boundary tests (Table~\ref{tab:boundary}).
\textbf{\cfg{Prompt-only}} generates boundary tests by prompting from the full discovery traces alone, without the workflow graph.
\textbf{\cfg{Random-turn}} extends \cfg{Prompt-only}. For each boundary test, it draws a turn uniformly at random from the traces and asks the test generator to perturb the workflow at that turn.
\textbf{\cfg{Graph-context}} uses the workflow graph that \tool builds but feeds it to the test generator only as a textual summary inside the prompt.
\textbf{\tool} gives the test generator the same traces and summary and, in addition, generates its tests at boundary targets enumerated from the structural locations of the workflow graph (Algorithm~\ref{alg:probe}).

\subsection{Benchmark Design}
\label{sec:eval-design}

\textbf{Privileged auditor.}
The benchmark uses an LLM \emph{auditor} to evaluate the test plan from discovery, and the test runs from execution.
In addition to the traces that the tester sees, the auditor has access to the agent's source code, policy, database, and tool calls, and to a \emph{functionality inventory} (a list of documented behaviors that the agent should support).

\textbf{Discovery-phase metrics.}
For $T$ generated tests with valid subset $T_v$, the \emph{validity rate} $|T_v|/|T|$ is the fraction the auditor accepts; it is reported for both functional and boundary tests under a kind-specific criterion: a functional test is valid if it is actionable and covers at least one inventory item; a boundary test if it is actionable and describes a real boundary (not an ordinary success path) whose success and failure criteria are decidable from the trace.
For both kinds, the auditor also checks that every record a test refers to by its identifier, such as a reservation code, exists in the agent’s database with the values the test gives for it, and that the test’s success and failure criteria agree with the policy; a test that fails either check is invalid.
For the valid functional tests, \emph{coverage recall} $|C|/|\mathcal{I}|$ is the fraction of the documented-behavior inventory $\mathcal{I}$ they cover.
For the valid boundary tests, there is no inventory of boundaries, so the auditor counts the distinct boundaries that these tests cover.
Specifically, it writes a \emph{boundary card} $(w,c,p,s)$ for each of these tests: the workflow~$w$ that the test's goal concerns; the condition~$c$, a rule of $w$ that the agent is expected to enforce (Section~\ref{sec:faultmodel}); the perturbation~$p$, a one-sentence description of the user action in the test at which the agent could violate $c$; and the test's success criteria~$s$.
The auditor then clusters the cards: two cards fall into one cluster only when they have the same $w$, $c$, and $s$ and the same kind of $p$, even if they differ in wording.
The number of clusters is the \emph{distinct-boundary count} $|U|$, and the \emph{duplicate rate} is $1-|U|/|P_v|$.

\textbf{Execution-phase metrics.}
For both functional and boundary tests, the auditor labels each valid test’s run as TP (the judge returned fail because of a fault of the agent), FP (the judge returned fail, but the agent was actually correct), TN (the judge returned pass, and the agent was correct), FN (the judge returned pass, although the agent showed a fault in the run), or inconclusive (INC: the run did not show the behavior that the test checks---the state that the test needs was never reached, the turn budget ran out, or a runtime error in the agent ended the run early; a run that the judge found inconclusive is also INC).
From these it reports the false-alarm rate $\mathrm{FAR}=\mathrm{FP}/(\mathrm{FP}+\mathrm{TN})$, accuracy $\mathrm{Acc}=(\mathrm{TP}+\mathrm{TN})/N$, and inconclusive rate $\mathrm{Inc}=\mathrm{INC}/N$ (with $N=\mathrm{TP}+\mathrm{FP}+\mathrm{TN}+\mathrm{FN}+\mathrm{INC}$).

\textbf{Seeded faults.}
Because the agents under test contain few real faults, in every domain we seed faults in two ways to evaluate fault detection effectiveness (RQ4).
Both ways use only generated tests labeled TN: the judge returned pass, and the auditor found this verdict correct.
We pick these tests in a random order and keep only those we manually confirm as valid.
First, we minimally edit one agent response in the trace of a kept test so that it states a wrong value or shows a boundary violation, such as bypassing confirmation.
The judge reads the edited trace directly and detects the fault when it returns fail.
Second, we seed faults into the agent to create \emph{mutant agents}.
We then evaluate whether the generated tests and the execution stage (which includes the judge) can kill the mutant agents.
This measures how effectively \tool detects faults in the agent, for instance in regression testing (Section~\ref{sec:overview}).
For a prompt bug, we modify one rule of the agent's policy that a test involves.
For a tool bug, we make one tool that a test uses return or write wrong data.
We design each seeded fault so that it is visible in the trace and relevant to the test's criteria.
On the mutant agent, we run the tests related to the fault.
If the judge returns fail, the mutant is killed.

\subsection{Setup}
\label{sec:eval-setup}

We evaluate \tool on the four text domains of $\tau^3$-bench~\cite{tau3bench}: Airline, Retail, and Telecom from $\tau^2$-bench~\cite{barres2025tau2bench}, and Banking from $\tau$-Knowledge~\cite{shi2026tauknowledge}.
We use these environments \emph{only} as black-box targets and do not use the official tasks, user simulators, or reward functions of $\tau^3$-bench.
We chose $\tau^3$-bench for three reasons.
First, it is widely used to evaluate LLM agents~\cite{cuadron2025saber} and models real-world customer-service domains~\cite{yao2024taubench}.
Second, its four domains cover diverse and complex workflows: bookings and orders in Airline and Retail, multi-step troubleshooting in Telecom, and knowledge retrieval with policy reasoning in Banking.
Third, its well-maintained policy documents~\cite{tau3bench} let us build a functionality inventory for coverage recall, and they describe the workflows and their boundaries clearly enough for us to manually check outputs such as the mined workflow graphs, the judge's verdicts, and the auditor's labels.
The inventory has $39$ items for Airline, $31$ for Retail, $30$ for Telecom, and $45$ for Banking ($145$ in total).

\textbf{Models.}
The tester makes all its LLM calls with MiMo-v2.6-pro: those of the discovery planner and driver, event abstraction, test generation, the runner, and the judge.
The agents under test run DeepSeek-v4-flash 0731.
We choose these two models because they offer strong performance at low cost.
The auditor produces most reported numbers, so our benchmark's correctness depends on it.
We therefore select the auditor model deliberately before the experiments, by measuring how closely each candidate model agrees with a human expert on coverage decisions.
We fix $26$ functional tests that \tool generated for the Airline domain and vary only the auditor model: each candidate reads the agent's source code and the inventory and decides which items the tests cover.
For each candidate, we compute the Jaccard overlap between its set of covered items and that of an expert (Table~\ref{tab:auditor}).
GLM-5.2 has the highest overlap ($0.94$) and the same coverage recall as the expert ($0.82$), so we choose it as the auditor.
All models run at temperature~$0$ to reduce the variation between runs.
We run MiMo-v2.6-pro and GLM-5.2 with reasoning disabled to save cost and latency, as enabling reasoning did not noticeably improve results in our experiments.

\textbf{Budgets.}
Discovery runs $R_{\mathrm{disc}}=12$ rounds, each one session of at most $B_{\mathrm{turn}}=6$ turns, starting with a warm-up of $m_{\mathrm{cap}}=1$ capability-discovery, $m_{\mathrm{hp}}=2$ happy-path, and $m_{\mathrm{cc}}=1$ consistency-check sessions.
Banking gets a larger discovery budget, $R_{\mathrm{disc}}=16$ rounds of at most $B_{\mathrm{turn}}=10$ turns, because its agent covers more services than the other three.
Test synthesis generates at most $N_f=50$ functional and $N_b=50$ boundary tests, with each test capped at $10$ turns.
We fixed these values to balance coverage against experimentation cost (LLM calls and wall-clock time).

\textbf{Repetitions.}
We repeat the experiments of RQ1--RQ5 five times to account for non-determinism and, unless noted otherwise, report the mean and standard deviation (SD) of each metric.

\begin{table}[tb]
\centering
\caption{Auditor-model selection on $26$ Airline functional tests}
\label{tab:auditor}
\begin{tabular}{@{}lrr@{}}
\toprule
\textbf{Auditor model} & \textbf{Recall} & \textbf{Overlap} \\
\midrule
Human reference  & 0.82 & --- \\
\textbf{GLM-5.2 (chosen)} & \textbf{0.82} & \textbf{0.94} \\
MiMo-v2.5-pro    & 0.90 & 0.91 \\
DeepSeek-v4-pro  & 0.90 & 0.91 \\
GPT-5.4-mini     & 0.85 & 0.86 \\
\bottomrule
\end{tabular}
\end{table}

\subsection{Results}
\label{sec:eval-results}

\textbf{RQ1: Functional testing effectiveness.}
The \tool columns of Table~\ref{tab:rq1} report the results for the functional tests in each domain.
The mean validity rate ranges from $0.87$ in Retail to $0.97$ in Airline.
The mean coverage recall is $0.95$ in Airline and Retail, and lower in Telecom ($0.87$) and Banking ($0.72$).
The mean FAR ranges from $0.01$ in Retail to $0.06$ in Telecom, and the mean Inc ranges from $0.05$ in Retail to $0.19$ in Banking.
Over the four domains and five repetitions, the auditor labeled $69$ runs of functional tests as TP.
We manually checked each run by comparing its trace and the agent's tool calls with the policy.
We confirmed that $43$ of the $69$ runs reveal an agent fault (column \textit {Conf.\ TP} gives the mean per repetition).
For example, functional tests found that the Retail agent stated a price difference of \$7.01 instead of \$8.01, and that the Telecom agent added data and charged \$4.00 before the user confirmed the price (Table~\ref{tab:faults}).

\rqanswer{RQ1}{\tool is effective at functional testing: in every domain, most of its functional tests are valid and together cover most of the documented behaviors. These test runs reveal real faults at a low false-alarm rate.}

\begin{table*}[tb]
\centering\footnotesize
\caption{Functional tests per domain and configuration (RQ1, RQ3)}
\label{tab:rq1}
\setlength{\tabcolsep}{4pt}
\begin{tabular}{@{}lrrrrrrrrr@{}}
\toprule
 & & \multicolumn{6}{c}{\textbf{\tool}} & \multicolumn{2}{c}{\textbf{\cfg{Naive}}} \\
\cmidrule(lr){3-8}\cmidrule(l){9-10}
\textbf{Domain} & $|\mathcal{I}|$ & \textbf{Valid} & \textbf{CovRec} & \textbf{Conf.\ TP} & \textbf{FAR} & \textbf{Acc} & \textbf{Inc} & \textbf{Valid} & \textbf{CovRec} \\
\midrule
Airline & 39 & 0.97$\pm$0.03 & 0.95$\pm$0.04 & 0.6$\pm$0.9 & 0.05$\pm$0.05 & 0.84$\pm$0.07 & 0.11$\pm$0.08 & 0.95$\pm$0.02 & 0.92$\pm$0.04 \\
Retail  & 31 & 0.87$\pm$0.06 & 0.95$\pm$0.03 & 1.2$\pm$1.3 & 0.01$\pm$0.01 & 0.94$\pm$0.05 & 0.05$\pm$0.04 & 0.89$\pm$0.05 & 0.98$\pm$0.02 \\
Telecom & 30 & 0.89$\pm$0.07 & 0.87$\pm$0.04 & 4.6$\pm$2.5 & 0.06$\pm$0.05 & 0.86$\pm$0.07 & 0.10$\pm$0.06 & 0.88$\pm$0.07 & 0.84$\pm$0.08 \\
Banking & 45 & 0.89$\pm$0.04 & 0.72$\pm$0.07 & 2.2$\pm$1.3 & 0.04$\pm$0.02 & 0.78$\pm$0.11 & 0.19$\pm$0.12 & 0.87$\pm$0.05 & 0.69$\pm$0.08 \\
\bottomrule
\end{tabular}
\end{table*}

\textbf{RQ2: Boundary testing effectiveness.}
The \tool rows of Table~\ref{tab:boundary} report the results for the boundary tests in each domain.
With at most $50$ boundary tests per domain and repetition, the mean distinct-boundary count ranges from $38.8$ in Telecom to $46.0$ in Airline.
The mean validity rate ranges from $0.89$ in Banking to $0.98$ in Airline, and the mean duplicate rate ranges from $0.05$ in Banking to $0.19$ in Telecom.
Telecom has the lowest count because it has the highest duplicate rate, not because it has few valid tests (mean validity rate $0.96$).
Banking has by far the largest workflow graphs ($95.8$ activities on average, compared with $30.0$--$42.8$ in the other domains), but its count ($42.4$) is below that of Airline ($46.0$).
Almost all valid tests in Banking cover different boundaries, but the count cannot exceed the number of valid tests.
Banking has the fewest valid tests.
The auditor rejects Banking boundary tests that assume a record or account state that the agent's database does not contain, or whose criteria do not agree with the policy.
The mean \textit {FAR} ranges from $0.03$ in Airline to $0.07$ in Banking.
Over the four domains and five repetitions, the auditor labeled $81$ runs of boundary tests as TP.
We manually confirmed that $43$ of them reveal an agent fault (column Conf.\ TP).
For example, boundary tests found that the Airline agent cancelled a reservation before the user confirmed, and that the Banking agent closed a credit-card account with a replacement card still in transit (Table~\ref{tab:faults}).

\begin{table*}[tb]
\centering\footnotesize
\caption{Real faults found in the agents under test}
\label{tab:faults}
\begin{tabularx}{\textwidth}{@{}ll>{\raggedright\arraybackslash}X>{\raggedright\arraybackslash}X@{}}
\toprule
\textbf{Domain} & \textbf{Test} & \textbf{What the test did} & \textbf{What the agent did wrong} \\
\midrule
Airline & Functional & Asked the cost of adding two checked bags to a two-passenger business reservation & Quoted \$100 for two bags that were within the free allowance of six bags \\
Airline & Boundary & Asked to cancel a two-passenger business round trip, giving a reason, then withdrew the request in the next turn & Cancelled the \$8,872 booking at once, without stating the action or asking for confirmation \\
Airline & Boundary & Asked to replace one flight of a two-flight economy trip & Priced the booked flights at current fares instead of the fares paid and quoted \$25 extra; after the user agreed, the change charged \$83 \\
Retail & Functional & Asked to change a \$94.01 kettle to a \$102.02 variant & Stated a price difference of \$7.01 instead of \$8.01, and reported \$7.01 paid after the gift card was charged \$8.01 \\
Retail & Boundary & Asked for another customer's order, insisting that it was the user's own & Refused, but revealed the owner's user ID and the city, state, and ZIP code of the shipping address \\
Retail & Boundary & Asked to cancel a pending order, gave a reason, then withdrew at the confirmation step & Cancelled the \$360.55 order before asking for confirmation, then said the cancellation could not be undone \\
Banking & Functional & Asked which personal checking account to open, then asked to open it & Recommended an account, saying it had no monthly fee; the fee is \$22.50, waived with a \$150 minimum daily balance \\
Banking & Boundary & Asked to close a credit-card account with a replacement card in transit & Closed the account without checking replacement orders; its own knowledge base search had returned the rule that blocks this closure \\
Banking & Boundary & Gave the wrong last four digits, expiration date, and CVV when activating a replacement debit card & Read out the stored last four digits, expiration date, and CVV \\
Telecom & Functional & Asked to resume a suspended line with a \$425 bill overdue & Resumed the line without checking bills; the policy allows this only after all overdue bills are paid \\
Telecom & Functional & Asked to add 2\,GB of data to a line and asked the price & Added the data and \$4.00 to the bill before answering; the policy requires confirming the price first \\
Telecom & Boundary & Asked to change an Issued bill to Overdue so that it could be paid & Sent a \$150 payment request for the Issued bill; the policy allows payment requests only for overdue bills \\
\bottomrule
\end{tabularx}
\end{table*}

\rqanswer{RQ2}{\tool is effective at boundary testing: in every domain, most of its boundary tests are valid and cover different boundaries. These runs reveal real faults at a low false-alarm rate.}

\textbf{RQ3: Ablation study.}
For discovery, Table~\ref{tab:rq1} compares \cfg{Naive} with \tool on the functional tests.
For \cfg{Naive}, which differs from Ours only in its discovery, the table gives only the discovery-stage metrics.
\cfg{Naive} and \tool use the same functional test generation, so differences in coverage recall arise mainly from the traces each discovery collected.
\cfg{Naive} has a lower mean validity rate and a lower mean coverage recall than Ours in Airline, Telecom, and Banking, and slightly higher ones in Retail.
For boundary test generation, Table~\ref{tab:boundary} compares \cfg{Prompt-only}, \cfg{Random-turn}, \cfg{Graph-context}, and \tool across domains.
On average, \cfg{Random-turn} covers the fewest distinct boundaries and has the highest duplicate rate in every domain because a randomly drawn turn does not necessarily have a boundary.
\cfg{Graph-context} covers on average $4.6$ more distinct boundaries than \cfg{Prompt-only} in Airline and $0.6$--$1.6$ fewer in the other three domains.
\tool covers the most distinct boundaries on average in every domain.
The margin over \cfg{Graph-context} ranges from $2.0$ in Telecom to $7.0$ in Retail.
The margin over \cfg{Prompt-only} ranges from $0.6$ in Telecom to $9.6$ in Airline, and the margin over \cfg{Random-turn} ranges from $6.2$ in Banking to $14.8$ in Airline.
\tool also has the lowest mean duplicate rate in Airline, Retail, and Banking.
In Telecom, \cfg{Prompt-only} and \cfg{Graph-context} have a lower mean duplicate rate ($0.17$) than Ours ($0.19$).
Within each domain, the mean validity rates of the four configurations differ by at most $0.09$, and their mean FARs by at most $0.05$.

\begin{table*}[tb]
\centering\footnotesize
\caption{Boundary tests per domain and configuration (RQ2, RQ3)}
\label{tab:boundary}
\setlength{\tabcolsep}{4pt}
\begin{tabular}{@{}llrrrrrrr@{}}
\toprule
\textbf{Domain} & \textbf{Configuration} & \textbf{Valid} & \textbf{Distinct} & \textbf{Dup.} & \textbf{Conf.\ TP} & \textbf{FAR} & \textbf{Acc} & \textbf{Inc} \\
\midrule
Airline & \cfg{Prompt-only}   & 0.98$\pm$0.01 & 36.4$\pm$3.8 & 0.26$\pm$0.07 & 1.2$\pm$0.8 & 0.02$\pm$0.02 & 0.89$\pm$0.02 & 0.09$\pm$0.03 \\
        & \cfg{Random-turn}   & 0.97$\pm$0.03 & 31.2$\pm$5.3 & 0.35$\pm$0.12 & 2.4$\pm$1.9 & 0.06$\pm$0.06 & 0.86$\pm$0.02 & 0.09$\pm$0.05 \\
        & \cfg{Graph-context} & 0.98$\pm$0.01 & 41.0$\pm$6.4 & 0.17$\pm$0.13 & 0.4$\pm$0.5 & 0.04$\pm$0.01 & 0.90$\pm$0.05 & 0.06$\pm$0.05 \\
        & \textbf{\tool} & 0.98$\pm$0.01 & 46.0$\pm$6.2 & 0.07$\pm$0.12 & 1.8$\pm$1.1 & 0.03$\pm$0.02 & 0.89$\pm$0.02 & 0.08$\pm$0.03 \\
\midrule
Retail  & \cfg{Prompt-only}   & 0.93$\pm$0.05 & 36.2$\pm$3.1 & 0.22$\pm$0.08 & 0.6$\pm$0.5 & 0.02$\pm$0.02 & 0.93$\pm$0.04 & 0.05$\pm$0.02 \\
        & \cfg{Random-turn}   & 0.94$\pm$0.01 & 31.6$\pm$3.2 & 0.33$\pm$0.07 & 0.0$\pm$0.0 & 0.04$\pm$0.01 & 0.89$\pm$0.04 & 0.07$\pm$0.04 \\
        & \cfg{Graph-context} & 0.92$\pm$0.05 & 35.6$\pm$6.7 & 0.23$\pm$0.10 & 0.4$\pm$0.9 & 0.02$\pm$0.02 & 0.96$\pm$0.02 & 0.02$\pm$0.02 \\
        & \textbf{\tool} & 0.95$\pm$0.03 & 42.6$\pm$6.2 & 0.10$\pm$0.14 & 1.2$\pm$0.8 & 0.04$\pm$0.02 & 0.91$\pm$0.04 & 0.05$\pm$0.05 \\
\midrule
Telecom & \cfg{Prompt-only}   & 0.92$\pm$0.05 & 38.2$\pm$5.8 & 0.17$\pm$0.10 & 2.6$\pm$1.5 & 0.09$\pm$0.07 & 0.87$\pm$0.07 & 0.05$\pm$0.03 \\
        & \cfg{Random-turn}   & 0.98$\pm$0.02 & 28.0$\pm$4.7 & 0.43$\pm$0.10 & 2.8$\pm$1.8 & 0.08$\pm$0.02 & 0.85$\pm$0.03 & 0.08$\pm$0.03 \\
        & \cfg{Graph-context} & 0.89$\pm$0.07 & 36.8$\pm$5.2 & 0.17$\pm$0.14 & 1.8$\pm$1.3 & 0.07$\pm$0.04 & 0.89$\pm$0.04 & 0.05$\pm$0.02 \\
        & \textbf{\tool} & 0.96$\pm$0.03 & 38.8$\pm$4.1 & 0.19$\pm$0.07 & 3.2$\pm$1.3 & 0.04$\pm$0.02 & 0.90$\pm$0.04 & 0.06$\pm$0.03 \\
\midrule
Banking & \cfg{Prompt-only}   & 0.84$\pm$0.10 & 39.0$\pm$4.1 & 0.07$\pm$0.05 & 2.0$\pm$1.2 & 0.02$\pm$0.02 & 0.86$\pm$0.04 & 0.12$\pm$0.02 \\
        & \cfg{Random-turn}   & 0.91$\pm$0.05 & 36.2$\pm$7.0 & 0.21$\pm$0.13 & 2.6$\pm$1.1 & 0.04$\pm$0.04 & 0.82$\pm$0.03 & 0.15$\pm$0.04 \\
        & \cfg{Graph-context} & 0.88$\pm$0.06 & 37.4$\pm$3.0 & 0.15$\pm$0.02 & 1.6$\pm$1.7 & 0.05$\pm$0.03 & 0.84$\pm$0.04 & 0.12$\pm$0.05 \\
        & \textbf{\tool} & 0.89$\pm$0.05 & 42.4$\pm$3.4 & 0.05$\pm$0.03 & 2.4$\pm$1.5 & 0.07$\pm$0.05 & 0.83$\pm$0.06 & 0.12$\pm$0.05 \\
\bottomrule
\end{tabular}
\end{table*}

\rqanswer{RQ3}{Both components contribute to the effectiveness of \tool: phase-guided discovery slightly raises the validity rate and coverage recall of the functional tests in three of the four domains. The boundary targets from the mined workflow graph increase the distinct-boundary count of the boundary tests in every domain and lower their duplicate rate in every domain except Telecom.}

\textbf{RQ4: Fault detection effectiveness.}
Table~\ref{tab:rq4} reports, per domain, in how many mutant traces the judge detected the fault and how many mutant agents of each kind were killed.
For the mutant traces, the table gives the mean $\pm$ SD over three readings of each trace by the judge.
On average, the judge detected the fault in $74.3$ of the $89$ mutant traces.
Per domain, the detection rate ranges from $0.80$ in Retail and Banking to $0.89$ in Airline.
For the mutant agents, the table gives the mean $\pm$ SD over five runs of each test.
On average, $28.0$ of the $32$ mutant agents with prompt bugs were killed.
Per domain, their kill rate ranges from $0.75$ in Banking to $0.98$ in Retail and Telecom.
On average, $17.0$ of the $29$ mutant agents with tool bugs were killed.
Per domain, their kill rate is at least $0.60$, except in Retail, where it is only $0.43$.
The kill rate in Retail is low for two reasons.
First, the agent often infers the correct value from the context instead of using an incorrect tool value, so the judge returns pass.
Second, the agent sometimes notices that a tool's result is wrong and transfers the user to a human representative, as its policy requires, so the judge returns inconclusive.
Both cases show a limitation of a judge that sees only the conversation traces.

\begin{table}[tb]
\centering\footnotesize
\caption{Faults seeded in traces and in agents (RQ4)}
\label{tab:rq4}
\setlength{\tabcolsep}{4pt}
\begin{tabular}{@{}lrrr@{}}
\toprule
 & \textbf{Mutant traces} & \multicolumn{2}{c}{\textbf{Mutant agents}} \\
 & (detected / seeded) & \multicolumn{2}{c}{(killed / seeded)} \\
\cmidrule(l){3-4}
\textbf{Domain} & & \textbf{Prompt bugs} & \textbf{Tool bugs} \\
\midrule
Airline & 19.7$\pm$0.6 / 22 & 6.4$\pm$0.9 / 8 & 4.8$\pm$0.8 / 8 \\
Retail  & 18.3$\pm$0.6 / 23 & 7.8$\pm$0.4 / 8 & 3.4$\pm$1.7 / 8 \\
Telecom & 18.7$\pm$0.6 / 22 & 7.8$\pm$0.4 / 8 & 5.6$\pm$0.9 / 8 \\
Banking & 17.7$\pm$0.6 / 22 & 6.0$\pm$1.4 / 8 & 3.2$\pm$0.4 / 5 \\
\bottomrule
\end{tabular}
\end{table}

\rqanswer{RQ4}{\tool is effective at detecting seeded faults: in every domain, the judge detects the fault in most mutant traces, and most mutant agents with prompt bugs are killed. Most mutant agents with tool bugs are also killed, except in Retail.}

\textbf{RQ5: Efficiency.}
Table~\ref{tab:cost} reports the mean number of LLM calls, tokens, and active wall-clock time of each configuration over the five repetitions.
The calls are those of the tester and of the agent under test.
The cost of each configuration includes a full discovery, test generation, and test execution.
For the functional tests, \tool takes, on average, from $0.74$\,h in Retail to $1.96$\,h in Banking, while \cfg{Naive} takes $0.03$ to $0.13$\,h longer.
For the boundary tests, \tool takes, on average, from $0.60$\,h in Retail to $1.80$\,h in Banking.
Within each domain, the mean times of the four boundary testing configurations differ by at most $0.15$\,h.

\begin{table*}[tb]
\centering\footnotesize
\caption{Mean cost per configuration over the five repetitions (RQ5)}
\label{tab:cost}
\setlength{\tabcolsep}{3.5pt}
\begin{tabular}{@{}lrrrrrrrrrrrr@{}}
\toprule
 & \multicolumn{3}{c}{\textbf{Airline}} & \multicolumn{3}{c}{\textbf{Retail}} & \multicolumn{3}{c}{\textbf{Telecom}} & \multicolumn{3}{c}{\textbf{Banking}} \\
\cmidrule(lr){2-4}\cmidrule(lr){5-7}\cmidrule(lr){8-10}\cmidrule(lr){11-13}
\textbf{Configuration} & Calls & Tokens (M) & Time (h) & Calls & Tokens (M) & Time (h) & Calls & Tokens (M) & Time (h) & Calls & Tokens (M) & Time (h) \\
\midrule
\multicolumn{13}{@{}l}{\textit{Functional testing}} \\
\cfg{Naive} & 1520 & 12.37 & 1.01 & 1096 & 7.24 & 0.77 & 1224 & 11.11 & 1.14 & 2251 & 84.56 & 2.02 \\
\tool & 1360 & 10.27 & 0.88 & 1166 & 7.26 & 0.74 & 1240 & 11.64 & 1.11 & 2169 & 89.41 & 1.96 \\
\midrule
\multicolumn{13}{@{}l}{\textit{Boundary testing}} \\
\cfg{Prompt-only} & 1361 & 9.86 & 0.86 & 1202 & 6.92 & 0.62 & 1150 & 9.69 & 0.94 & 2012 & 67.17 & 1.67 \\
\cfg{Random-turn} & 1432 & 10.89 & 0.87 & 1121 & 6.54 & 0.59 & 1154 & 9.94 & 0.96 & 2175 & 86.12 & 1.82 \\
\cfg{Graph-context} & 1344 & 9.53 & 0.86 & 1184 & 6.86 & 0.61 & 1225 & 10.22 & 0.98 & 2125 & 74.06 & 1.80 \\
\tool & 1342 & 9.82 & 0.86 & 1113 & 6.54 & 0.60 & 1166 & 9.88 & 0.97 & 2058 & 76.38 & 1.80 \\
\bottomrule
\end{tabular}
\end{table*}

\rqanswer{RQ5}{\tool runs at an acceptable cost: in every domain, functional testing and boundary testing each take less than two hours on average, including a full discovery. Mining the workflow graph to guide boundary testing does not noticeably increase the cost compared with generating boundary tests by prompting from the discovery traces alone.}

\subsection{Threats to Validity}
\label{sec:eval-threats}

\textbf{Construct.}
Every metric comes from the LLM auditor, so an unreliable auditor would corrupt all of them.
We run it on a different model family from the tester (GLM-5.2 \vs MiMo-v2.6-pro) to avoid self-audit bias, and validate only its covered functionality set against an expert reference on the airline domain (Section~\ref{sec:eval-setup}).
The distinct-boundary count depends on how the auditor clusters the boundary cards.
Because the auditor puts two cards into one cluster only when they clearly describe the same boundary, it may count two similar boundaries as two.
We leave a more accurate clustering of the boundary cards to future work.

\textbf{Internal.}
Results for each configuration vary across repetitions, as the SDs in the tables show.
A difference between two configurations that is smaller than this variation may therefore not be real.
The judge of \tool decides from the trace alone, as a deployed black-box tester must, so faults that leave the trace unchanged remain undetectable by construction.
Therefore, we seed only faults that can change the trace.
A fault that alters backend state without changing the agent's responses, such as a tool that silently changes a refund amount, remains undetectable from the conversation.
The values that a functional test records come from the agent's responses during discovery (Section~\ref{sec:synthesis}).
If one of those responses stated a wrong value, the test treats that value as expected.

\textbf{External.}
We study four $\tau^3$-bench text domains, all of which are tool-using service agents driven by a single model family; other architectures, modalities, non-service domains, and LLMs may differ.
\tool also assumes a recurring interaction structure (routes, slots, validation, confirmation gates) that is observable in bounded traces; for highly open-ended or non-repeatable agents, the mined graph may remain thin, yielding fewer boundary targets than a trace-only baseline.

\section{Related Work}
\label{sec:related}

\textbf{Evaluating and testing LLM agents.}
Agent benchmarks rank model capability on fixed task suites with programmatic or simulated-user scoring~\cite{yao2024taubench,zhou2024webarena,mialon2023gaia,trivedi2024appworld,jimenez2024swebench,qin2024toolllm}.
Closer to testing, TRACE evolves benchmark tasks under a validate-by-reproduce oracle~\cite{trace2025}, TestAgent sequences follow-up probes adaptively~\cite{testagent2024}, and Graph2Eval seeds agent tasks from an external knowledge graph~\cite{chen2025graph2eval}; LLMs have also been used to reverse-engineer black-box systems~\cite{geng2025reverseeng}.
These explore or adapt tasks, but none first induces a behavioral model of the agent under test and then tests systematically against it.
\tool does both: it discovers the agent by interaction, synthesizes tests from the mined model, and judges each run with an oracle that reads only the conversation.

\textbf{Mining behavioral models.}
\tool's graph is a mined behavioral model.
Specification mining infers models from execution traces~\cite{ammons2002mining,cook1998discovering}, including state-machine inference such as GK-tail and CSight~\cite{lorenzoli2008gktail,beschastnikh2014csight}; process mining discovers process models from event logs~\cite{vanderaalst2016processmining,vanderaalst2004workflowmining,leemans2013inductive}, with directly-follows graphs the pragmatic workhorse~\cite{vanderaalst2019limitations}---which can imply paths never observed as full traces---on which we build, alongside event abstraction for low-level logs~\cite{tax2016eventabstraction,vanzelst2021eventabstraction}.
The name \emph{workflow graph} also denotes a design-time business-process formalism~\cite{sadiq1999workflow}; ours has the opposite provenance, induced from the running system's traces.
Active automata learning reconstructs machines through membership and equivalence queries~\cite{angluin1987learning} that a conversational target does not answer, while dialog-structure induction and user simulators build models from conversations for analysis, training, or evaluation~\cite{shi2019dialogstructure,burdisso2024dialog2flow,schatzmann2007agenda,zhu2020convlab2}.
All of these mine a model to \emph{describe} behavior; \tool mines one to \emph{act on} it: generating tests, replaying routes to target states, and stressing them under a tester that chooses what to observe next.

\textbf{Model-based and stateful API testing.}
Model-based testing derives tests from behavioral models~\cite{utting2012taxonomy}, but those models are typically authored and serve as the test oracle; ours is mined from the system under test, so it can guide test generation but cannot judge correctness itself.
Our prerequisite routes correspond to the transfer sequences, or preambles, that FSM conformance testing uses to drive a stateful system into the state under test~\cite{lee1996fsm,chow1978fsm}.
Stateful REST API testing infers operation dependencies to build executable sequences: RESTler from specifications~\cite{atlidakis2019restler}, EvoMaster by search~\cite{arcuri2019evomaster}, Morest with an execution-refined property graph~\cite{liu2022morest}, and KAT with LLM-built dependency graphs~\cite{le2024kat}; see Golmohammadi et al.\ for a survey~\cite{zhang2023restapis}.
The conversational workflow graph serves as the dependency model, but is mined from natural-language conversations, and its routes replay user actions rather than API calls.

\textbf{LLM test generation, oracles, and boundary testing.}
LLM-based test generators validate candidates against executable signals---compilation, coverage, a crashing oracle---and repair the rest: TestPilot~\cite{schafer2024testpilot}, CoverUp~\cite{pizzorno2025coverup}, ChatTester~\cite{yuan2024chattester}, CodaMosa~\cite{lemieux2023codamosa}, and TestART~\cite{gu2024testart}.
A black-box conversational agent exposes none of these signals, so the oracle problem~\cite{barr2015oracle} is acute: classical pseudo-oracles need an independent implementation or checker~\cite{davis1981pseudo,weyuker1982nontestable}, LLM-as-judge is biased~\cite{zheng2023judging}, and intrinsic self-correction is unreliable without external grounding~\cite{huang2024selfcorrect,stechly2024selfverification,madaan2023selfrefine,gou2024critic}.
\tool's answer is to separate the actor from the oracle and leave every verdict to a judge confined to the visible trace---the weak black-box analogue of a pseudo-oracle.
For the perturbations themselves, boundary value analysis targets the edges of input domains~\cite{myers2011artoftesting,dobslaw2020boundary}, robustness testing perturbs single calls while excluding state~\cite{kropp1998ballista}, and metamorphic testing relates outputs when no oracle exists~\cite{chen2018metamorphic}, while prompt-injection studies motivate one family of perturbation~\cite{greshake2023indirect,perez2022ignore}; \tool lifts the boundary idea from input edges to workflow-state preconditions, which is precisely where conversational agents fail.

\section{Conclusion}
\label{sec:conclusion}
We present \tool, a black-box testing framework for conversational LLM agents that targets state-dependent workflow failures.
Such failures occur only after the conversation has reached a particular state of a workflow, which often takes several turns.
\tool therefore mines a \emph{conversational workflow graph} from observed interactions and generates boundary tests that replay the conversational path to a workflow boundary and then perturb one user action there.
It also generates functional tests that replay successful workflows observed during discovery.
On four $\tau^3$-bench agents, most tests of both kinds are valid, and the boundary tests cover more distinct boundaries than tests generated without the workflow graph.
Both kinds of tests reveal real faults at a low false-alarm rate, and \tool detects most seeded faults.
\tool and our benchmark can serve as a foundation for future research.

\section*{Data Availability}
The code, prompts, and data used in this paper will be made available upon acceptance.

\section*{Acknowledgements}
This work has emanated from research jointly funded by Taighde Éireann – Research Ireland under Grant Number. 13/RC/2094\_2, and by Genesys Cloud Services, Inc.

\bibliographystyle{IEEEtran}
\bibliography{references}

@article{yao2024taubench,
  author  = {Yao, Shunyu and Shinn, Noah and Razavi, Pedram and Narasimhan, Karthik},
  title   = {{$\tau$}-bench: A Benchmark for Tool-Agent-User Interaction in Real-World Domains},
  journal = {arXiv preprint arXiv:2406.12045},
  year    = {2024},
}

@article{barres2025tau2bench,
  author  = {Barres, Victor and Dong, Honghua and Ray, Soham and Si, Xujie and Narasimhan, Karthik},
  title   = {{$\tau^2$}-Bench: Evaluating Conversational Agents in a Dual-Control Environment},
  journal = {arXiv preprint arXiv:2506.07982},
  year    = {2025},
}

@article{shi2026tauknowledge,
  author  = {Shi, Quan and Zytek, Alexandra and Razavi, Pedram and Narasimhan, Karthik and Barres, Victor},
  title   = {{$\tau$}-Knowledge: Evaluating Conversational Agents over Unstructured Knowledge},
  journal = {arXiv preprint arXiv:2603.04370},
  year    = {2026},
}

@misc{tau3bench,
  author       = {{Sierra Research}},
  title        = {{$\tau^3$}-Bench: From Text-Only to Multimodal, Knowledge-Aware Agent Evaluation},
  year         = {2026},
  howpublished = {\url{https://github.com/sierra-research/tau2-bench}},
  note         = {Accessed: 2026-06-30},
}

@article{sorokin2026stellar,
  author  = {Sorokin, Lev and Vasilev, Ivan and Friedl, Ken E. and Stocco, Andrea},
  title   = {{STELLAR}: A Search-Based Testing Framework for Large Language Model Applications},
  journal = {arXiv preprint arXiv:2601.00497},
  year    = {2026},
}

@inproceedings{ahmed2026specops,
  author    = {Ahmed, Syed Yusuf and Feng, Shiwei and Bae, Chanwoo and Barrus, Calix and Zhang, Xiangyu},
  title     = {{SpecOps}: A Fully Automated {AI} Agent Testing Framework in Real-World {GUI} Environments},
  booktitle = {Proceedings of the 48th {IEEE/ACM} International Conference on Software Engineering ({ICSE})},
  year      = {2026},
}

@inproceedings{zhou2024webarena,
  author    = {Zhou, Shuyan and Xu, Frank F. and Zhu, Hao and Zhou, Xuhui and Lo, Robert and Sridhar, Abishek and Cheng, Xianyi and Ou, Tianyue and Bisk, Yonatan and Fried, Daniel and Alon, Uri and Neubig, Graham},
  title     = {{WebArena}: A Realistic Web Environment for Building Autonomous Agents},
  booktitle = {International Conference on Learning Representations ({ICLR})},
  year      = {2024},
}

@inproceedings{jimenez2024swebench,
  author    = {Jimenez, Carlos E. and Yang, John and Wettig, Alexander and Yao, Shunyu and Pei, Kexin and Press, Ofir and Narasimhan, Karthik},
  title     = {{SWE}-bench: Can Language Models Resolve Real-World {GitHub} Issues?},
  booktitle = {International Conference on Learning Representations ({ICLR})},
  year      = {2024},
}

@inproceedings{mialon2023gaia,
  author    = {Mialon, Gr{\'e}goire and Fourrier, Cl{\'e}mentine and Swift, Craig and Wolf, Thomas and LeCun, Yann and Scialom, Thomas},
  title     = {{GAIA}: A Benchmark for General {AI} Assistants},
  booktitle = {International Conference on Learning Representations ({ICLR})},
  year      = {2024},
}

@inproceedings{trivedi2024appworld,
  author    = {Trivedi, Harsh and Khot, Tushar and Hartmann, Mareike and Manku, Ruskin and Dong, Vinty and Li, Edward and Gupta, Shashank and Sabharwal, Ashish and Balasubramanian, Niranjan},
  title     = {{AppWorld}: A Controllable World of Apps and People for Benchmarking Interactive Coding Agents},
  booktitle = {Annual Meeting of the Association for Computational Linguistics ({ACL})},
  pages     = {16022--16076},
  year      = {2024},
}

@inproceedings{qin2024toolllm,
  author    = {Qin, Yujia and Liang, Shihao and Ye, Yining and Zhu, Kunlun and Yan, Lan and Lu, Yaxi and Lin, Yankai and Cong, Xin and Tang, Xiangru and Qian, Bill and Zhao, Sihan and Hong, Lauren and Tian, Runchu and Xie, Ruobing and Zhou, Jie and Gerstein, Mark and Li, Dahai and Liu, Zhiyuan and Sun, Maosong},
  title     = {{ToolLLM}: Facilitating Large Language Models to Master 16000+ Real-World {APIs}},
  booktitle = {International Conference on Learning Representations ({ICLR})},
  year      = {2024},
}

@article{trace2025,
  author  = {Guo, Dadi and Zhou, Tianyi and Liu, Dongrui and Qian, Chen and Ren, Qihan and Shao, Shuai and Fan, Zhiyuan and Fung, Yi R. and Wang, Kun and Zhang, Linfeng and Shao, Jing},
  title   = {Towards Self-Evolving Benchmarks: Synthesizing Agent Trajectories via Test-Time Exploration under Validate-by-Reproduce Paradigm},
  journal = {arXiv preprint arXiv:2510.00415},
  year    = {2025},
}

@article{testagent2024,
  author  = {Wang, Wanying and Ma, Zeyu and Wang, Xuhong and Zhang, Yangchun and Liu, Pengfei and Chen, Mingang},
  title   = {{TestAgent}: Automatic Benchmarking and Exploratory Interaction for Evaluating {LLMs} in Vertical Domains},
  journal = {arXiv preprint arXiv:2410.11507},
  year    = {2024},
}

@article{chen2025graph2eval,
  author  = {Chen, Yurun and Hu, Xavier and Liu, Yuhan and Wang, Ziqi and Liao, Zeyi and Chen, Lin and Wei, Feng and Qian, Yuxi and Zheng, Bo and Yin, Keting and Zhang, Shengyu},
  title   = {{Graph2Eval}: Automatic Multimodal Task Generation for Agents via Knowledge Graphs},
  journal = {arXiv preprint arXiv:2510.00507},
  year    = {2025},
}

@article{geng2025reverseeng,
  author  = {Geng, Jiayi and Chen, Howard and Arumugam, Dilip and Griffiths, Thomas L.},
  title   = {Are Large Language Models Reliable {AI} Scientists? Assessing Reverse-Engineering of Black-Box Systems},
  journal = {arXiv preprint arXiv:2505.17968},
  year    = {2025},
}

@inproceedings{mcgregor2021incidents,
  author    = {McGregor, Sean},
  title     = {Preventing Repeated Real World {AI} Failures by Cataloging Incidents: The {AI} Incident Database},
  booktitle = {Proceedings of the {AAAI} Conference on Artificial Intelligence ({IAAI} track)},
  volume    = {35},
  number    = {17},
  pages     = {15458--15463},
  year      = {2021},
}

@inproceedings{lejeune2025realharm,
  author    = {Le Jeune, Pierre and Liu, Jiaen and Rossi, Luca and Dora, Matteo},
  title     = {{RealHarm}: A Collection of Real-World Language Model Application Failures},
  booktitle = {Proceedings of the First Workshop on Large Language Model Security ({LLMSEC})},
  pages     = {87--100},
  year      = {2025},
}

@inproceedings{yao2023react,
  author    = {Yao, Shunyu and Zhao, Jeffrey and Yu, Dian and Du, Nan and Shafran, Izhak and Narasimhan, Karthik and Cao, Yuan},
  title     = {{ReAct}: Synergizing Reasoning and Acting in Language Models},
  booktitle = {International Conference on Learning Representations ({ICLR})},
  year      = {2023},
}

@inproceedings{schick2023toolformer,
  author    = {Schick, Timo and Dwivedi-Yu, Jane and Dess{\`i}, Roberto and Raileanu, Roberta and Lomeli, Maria and Hambro, Eric and Zettlemoyer, Luke and Cancedda, Nicola and Scialom, Thomas},
  title     = {Toolformer: Language Models Can Teach Themselves to Use Tools},
  booktitle = {Advances in Neural Information Processing Systems ({NeurIPS})},
  year      = {2023},
}

@book{vanderaalst2016processmining,
  author    = {van der Aalst, Wil M. P.},
  title     = {Process Mining: Data Science in Action},
  edition   = {2},
  publisher = {Springer},
  year      = {2016},
}

@article{vanderaalst2004workflowmining,
  author  = {van der Aalst, Wil M. P. and Weijters, Ton and Maruster, Laura},
  title   = {Workflow Mining: Discovering Process Models from Event Logs},
  journal = {IEEE Transactions on Knowledge and Data Engineering},
  volume  = {16},
  number  = {9},
  pages   = {1128--1142},
  year    = {2004},
}

@inproceedings{vanderaalst2019limitations,
  author    = {van der Aalst, Wil M. P.},
  title     = {A Practitioner's Guide to Process Mining: Limitations of the Directly-Follows Graph},
  booktitle = {International Conference on Enterprise Information Systems ({CENTERIS}), Procedia Computer Science vol.~164},
  pages     = {321--328},
  year      = {2019},
}

@inproceedings{leemans2013inductive,
  author    = {Leemans, Sander J. J. and Fahland, Dirk and van der Aalst, Wil M. P.},
  title     = {Discovering Block-Structured Process Models from Event Logs---A Constructive Approach},
  booktitle = {Application and Theory of Petri Nets and Concurrency, {LNCS} 7927},
  pages     = {311--329},
  year      = {2013},
}

@inproceedings{tax2016eventabstraction,
  author    = {Tax, Niek and Sidorova, Natalia and Haakma, Reinder and van der Aalst, Wil M. P.},
  title     = {Event Abstraction for Process Mining Using Supervised Learning Techniques},
  booktitle = {Intelligent Systems Conference ({IntelliSys})},
  pages     = {251--269},
  year      = {2016},
}

@article{vanzelst2021eventabstraction,
  author  = {van Zelst, Sebastiaan J. and Mannhardt, Felix and de Leoni, Massimiliano and Koschmider, Anne},
  title   = {Event Abstraction in Process Mining: Literature Review and Taxonomy},
  journal = {Granular Computing},
  volume  = {6},
  number  = {3},
  pages   = {719--736},
  year    = {2021},
}

@inproceedings{sadiq1999workflow,
  author    = {Sadiq, Wasim and Orlowska, Maria E.},
  title     = {Applying Graph Reduction Techniques for Identifying Structural Conflicts in Process Models},
  booktitle = {International Conference on Advanced Information Systems Engineering ({CAiSE}), {LNCS} 1626},
  pages     = {195--209},
  year      = {1999},
}

@inproceedings{ammons2002mining,
  author    = {Ammons, Glenn and Bod{\'i}k, Rastislav and Larus, James R.},
  title     = {Mining Specifications},
  booktitle = {ACM SIGPLAN-SIGACT Symposium on Principles of Programming Languages ({POPL})},
  pages     = {4--16},
  year      = {2002},
}

@article{cook1998discovering,
  author  = {Cook, Jonathan E. and Wolf, Alexander L.},
  title   = {Discovering Models of Software Processes from Event-Based Data},
  journal = {ACM Transactions on Software Engineering and Methodology},
  volume  = {7},
  number  = {3},
  pages   = {215--249},
  year    = {1998},
}

@inproceedings{lorenzoli2008gktail,
  author    = {Lorenzoli, Davide and Mariani, Leonardo and Pezz{\`e}, Mauro},
  title     = {Automatic Generation of Software Behavioral Models},
  booktitle = {International Conference on Software Engineering ({ICSE})},
  pages     = {501--510},
  year      = {2008},
}

@inproceedings{beschastnikh2014csight,
  author    = {Beschastnikh, Ivan and Brun, Yuriy and Ernst, Michael D. and Krishnamurthy, Arvind},
  title     = {Inferring Models of Concurrent Systems from Logs of Their Behavior with {CSight}},
  booktitle = {International Conference on Software Engineering ({ICSE})},
  pages     = {468--479},
  year      = {2014},
}

@article{angluin1987learning,
  author  = {Angluin, Dana},
  title   = {Learning Regular Sets from Queries and Counterexamples},
  journal = {Information and Computation},
  volume  = {75},
  number  = {2},
  pages   = {87--106},
  year    = {1987},
}

@inproceedings{shi2019dialogstructure,
  author    = {Shi, Weiyan and Zhao, Tiancheng and Yu, Zhou},
  title     = {Unsupervised Dialog Structure Learning},
  booktitle = {Conference of the North American Chapter of the Association for Computational Linguistics ({NAACL-HLT})},
  pages     = {1797--1807},
  year      = {2019},
}

@inproceedings{burdisso2024dialog2flow,
  author    = {Burdisso, Sergio and Madikeri, Srikanth and Motlicek, Petr},
  title     = {{Dialog2Flow}: Pre-training Soft-Contrastive Action-Driven Sentence Embeddings for Automatic Dialog Flow Extraction},
  booktitle = {Conference on Empirical Methods in Natural Language Processing ({EMNLP})},
  pages     = {5421--5440},
  year      = {2024},
}

@inproceedings{schatzmann2007agenda,
  author    = {Schatzmann, Jost and Thomson, Blaise and Weilhammer, Karl and Ye, Hui and Young, Steve},
  title     = {Agenda-Based User Simulation for Bootstrapping a {POMDP} Dialogue System},
  booktitle = {Human Language Technologies: Conference of the North American Chapter of the Association for Computational Linguistics ({NAACL-HLT}), Short Papers},
  pages     = {149--152},
  year      = {2007},
}

@inproceedings{zhu2020convlab2,
  author    = {Zhu, Qi and Zhang, Zheng and Fang, Yan and Li, Xiang and Takanobu, Ryuichi and Li, Jinchao and Peng, Baolin and Gao, Jianfeng and Zhu, Xiaoyan and Huang, Minlie},
  title     = {{ConvLab-2}: An Open-Source Toolkit for Building, Evaluating, and Diagnosing Dialogue Systems},
  booktitle = {Annual Meeting of the Association for Computational Linguistics ({ACL}), System Demonstrations},
  pages     = {142--149},
  year      = {2020},
}

@article{barr2015oracle,
  author  = {Barr, Earl T. and Harman, Mark and McMinn, Phil and Shahbaz, Muzammil and Yoo, Shin},
  title   = {The Oracle Problem in Software Testing: A Survey},
  journal = {IEEE Transactions on Software Engineering},
  volume  = {41},
  number  = {5},
  pages   = {507--525},
  year    = {2015},
}

@inproceedings{davis1981pseudo,
  author    = {Davis, Martin D. and Weyuker, Elaine J.},
  title     = {Pseudo-Oracles for Non-Testable Programs},
  booktitle = {Proceedings of the {ACM} '81 Conference},
  pages     = {254--257},
  year      = {1981},
}

@article{weyuker1982nontestable,
  author  = {Weyuker, Elaine J.},
  title   = {On Testing Non-Testable Programs},
  journal = {The Computer Journal},
  volume  = {25},
  number  = {4},
  pages   = {465--470},
  year    = {1982},
}

@article{lee1996fsm,
  author  = {Lee, David and Yannakakis, Mihalis},
  title   = {Principles and Methods of Testing Finite State Machines---A Survey},
  journal = {Proceedings of the IEEE},
  volume  = {84},
  number  = {8},
  pages   = {1090--1123},
  year    = {1996},
}

@article{chow1978fsm,
  author  = {Chow, Tsun S.},
  title   = {Testing Software Design Modeled by Finite-State Machines},
  journal = {IEEE Transactions on Software Engineering},
  volume  = {SE-4},
  number  = {3},
  pages   = {178--187},
  year    = {1978},
}

@inproceedings{kropp1998ballista,
  author    = {Kropp, Nathan P. and Koopman, Philip J. and Siewiorek, Daniel P.},
  title     = {Automated Robustness Testing of Off-the-Shelf Software Components},
  booktitle = {International Symposium on Fault-Tolerant Computing ({FTCS})},
  pages     = {230--239},
  year      = {1998},
}

@book{myers2011artoftesting,
  author    = {Myers, Glenford J. and Sandler, Corey and Badgett, Tom},
  title     = {The Art of Software Testing},
  edition   = {3},
  publisher = {Wiley},
  year      = {2011},
}

@inproceedings{dobslaw2020boundary,
  author    = {Dobslaw, Felix and de Oliveira Neto, Francisco Gomes and Feldt, Robert},
  title     = {Boundary Value Exploration for Software Analysis},
  booktitle = {IEEE International Conference on Software Testing, Verification and Validation Workshops ({ICSTW})},
  pages     = {346--353},
  year      = {2020},
}

@article{chen2018metamorphic,
  author  = {Chen, Tsong Yueh and Kuo, Fei-Ching and Liu, Huai and Poon, Pak-Lok and Towey, Dave and Tse, T. H. and Zhou, Zhi Quan},
  title   = {Metamorphic Testing: A Review of Challenges and Opportunities},
  journal = {ACM Computing Surveys},
  volume  = {51},
  number  = {1},
  pages   = {4:1--4:27},
  year    = {2018},
}

@article{utting2012taxonomy,
  author  = {Utting, Mark and Pretschner, Alexander and Legeard, Bruno},
  title   = {A Taxonomy of Model-Based Testing Approaches},
  journal = {Software Testing, Verification and Reliability},
  volume  = {22},
  number  = {5},
  pages   = {297--312},
  year    = {2012},
}

@inproceedings{atlidakis2019restler,
  author    = {Atlidakis, Vaggelis and Godefroid, Patrice and Polishchuk, Marina},
  title     = {{RESTler}: Stateful {REST} {API} Fuzzing},
  booktitle = {International Conference on Software Engineering ({ICSE})},
  pages     = {748--758},
  year      = {2019},
}

@article{arcuri2019evomaster,
  author  = {Arcuri, Andrea},
  title   = {{RESTful} {API} Automated Test Case Generation with {EvoMaster}},
  journal = {ACM Transactions on Software Engineering and Methodology},
  volume  = {28},
  number  = {1},
  pages   = {3:1--3:37},
  year    = {2019},
}

@inproceedings{liu2022morest,
  author    = {Liu, Yi and Li, Yuekang and Deng, Gelei and Liu, Yang and Wan, Ruiyuan and Wu, Runchao and Ji, Dandan and Xu, Shiheng and Bao, Minli},
  title     = {Morest: Model-Based {RESTful} {API} Testing with Execution Feedback},
  booktitle = {International Conference on Software Engineering ({ICSE})},
  pages     = {1406--1417},
  year      = {2022},
}

@inproceedings{le2024kat,
  author    = {Le, Tri and Tran, Thien and Cao, Duy and Le, Vy and Nguyen, Tien N. and Nguyen, Vu},
  title     = {{KAT}: Dependency-Aware Automated {API} Testing with Large Language Models},
  booktitle = {IEEE Conference on Software Testing, Verification and Validation ({ICST})},
  year      = {2024},
}

@article{zhang2023restapis,
  author  = {Golmohammadi, Amid and Zhang, Man and Arcuri, Andrea},
  title   = {Testing {RESTful} {APIs}: A Survey},
  journal = {ACM Transactions on Software Engineering and Methodology},
  volume  = {33},
  number  = {1},
  pages   = {27:1--27:41},
  year    = {2023},
}

@article{schafer2024testpilot,
  author  = {Sch{\"a}fer, Max and Nadi, Sarah and Eghbali, Aryaz and Tip, Frank},
  title   = {An Empirical Evaluation of Using Large Language Models for Automated Unit Test Generation},
  journal = {IEEE Transactions on Software Engineering},
  volume  = {50},
  number  = {1},
  pages   = {85--105},
  year    = {2024},
}

@article{pizzorno2025coverup,
  author  = {Altmayer Pizzorno, Juan and Berger, Emery D.},
  title   = {{CoverUp}: Effective High Coverage Test Generation for {Python}},
  journal = {Proceedings of the ACM on Software Engineering},
  volume  = {2},
  number  = {FSE},
  year    = {2025},
}

@inproceedings{yuan2024chattester,
  author    = {Yuan, Zhiqiang and Liu, Mingwei and Ding, Shiji and Wang, Kaixin and Chen, Yixuan and Peng, Xin and Lou, Yiling},
  title     = {Evaluating and Improving {ChatGPT} for Unit Test Generation},
  booktitle = {ACM International Conference on the Foundations of Software Engineering ({FSE})},
  pages     = {1703--1726},
  year      = {2024},
}

@inproceedings{lemieux2023codamosa,
  author    = {Lemieux, Caroline and Inala, Jeevana Priya and Lahiri, Shuvendu K. and Sen, Siddhartha},
  title     = {{CodaMosa}: Escaping Coverage Plateaus in Test Generation with Pre-trained Large Language Models},
  booktitle = {International Conference on Software Engineering ({ICSE})},
  pages     = {919--931},
  year      = {2023},
}

@article{gu2024testart,
  author  = {Gu, Siqi and Zhang, Quanjun and Li, Kecheng and Fang, Chunrong and Tian, Fangyuan and Zhu, Liuchuan and Zhou, Jianyi and Chen, Zhenyu},
  title   = {{TestART}: Improving {LLM}-Based Unit Testing via Co-Evolution of Automated Generation and Repair Iteration},
  journal = {arXiv preprint arXiv:2408.03095},
  year    = {2024},
}

@inproceedings{huang2024selfcorrect,
  author    = {Huang, Jie and Chen, Xinyun and Mishra, Swaroop and Zheng, Huaixiu Steven and Yu, Adams Wei and Song, Xinying and Zhou, Denny},
  title     = {Large Language Models Cannot Self-Correct Reasoning Yet},
  booktitle = {International Conference on Learning Representations ({ICLR})},
  year      = {2024},
}

@article{stechly2024selfverification,
  author  = {Stechly, Kaya and Valmeekam, Karthik and Kambhampati, Subbarao},
  title   = {On the Self-Verification Limitations of Large Language Models on Reasoning and Planning Tasks},
  journal = {arXiv preprint arXiv:2402.08115},
  year    = {2024},
}

@inproceedings{gou2024critic,
  author    = {Gou, Zhibin and Shao, Zhihong and Gong, Yeyun and Shen, Yelong and Yang, Yujiu and Duan, Nan and Chen, Weizhu},
  title     = {{CRITIC}: Large Language Models Can Self-Correct with Tool-Interactive Critiquing},
  booktitle = {International Conference on Learning Representations ({ICLR})},
  year      = {2024},
}

@inproceedings{madaan2023selfrefine,
  author    = {Madaan, Aman and Tandon, Niket and Gupta, Prakhar and Hallinan, Skyler and Gao, Luyu and Wiegreffe, Sarah and Alon, Uri and Dziri, Nouha and Prabhumoye, Shrimai and Yang, Yiming and Gupta, Shashank and Majumder, Bodhisattwa Prasad and Hermann, Katherine and Welleck, Sean and Yazdanbakhsh, Amir and Clark, Peter},
  title     = {{Self-Refine}: Iterative Refinement with Self-Feedback},
  booktitle = {Advances in Neural Information Processing Systems ({NeurIPS})},
  year      = {2023},
}

@inproceedings{zheng2023judging,
  author    = {Zheng, Lianmin and Chiang, Wei-Lin and Sheng, Ying and Zhuang, Siyuan and Wu, Zhanghao and Zhuang, Yonghao and Lin, Zi and Li, Zhuohan and Li, Dacheng and Xing, Eric P. and Zhang, Hao and Gonzalez, Joseph E. and Stoica, Ion},
  title     = {Judging {LLM-as-a-Judge} with {MT-Bench} and {Chatbot Arena}},
  booktitle = {Advances in Neural Information Processing Systems ({NeurIPS}), Datasets and Benchmarks Track},
  year      = {2023},
}

@inproceedings{greshake2023indirect,
  author    = {Greshake, Kai and Abdelnabi, Sahar and Mishra, Shailesh and Endres, Christoph and Holz, Thorsten and Fritz, Mario},
  title     = {Not What You've Signed Up For: Compromising Real-World {LLM}-Integrated Applications with Indirect Prompt Injection},
  booktitle = {ACM Workshop on Artificial Intelligence and Security ({AISec})},
  pages     = {79--90},
  year      = {2023},
}

@inproceedings{perez2022ignore,
  author    = {Perez, F{\'a}bio and Ribeiro, Ian},
  title     = {Ignore Previous Prompt: Attack Techniques for Language Models},
  booktitle = {NeurIPS ML Safety Workshop},
  year      = {2022},
}

@inproceedings{nakash2025redteaming,
  author    = {Nakash, Itay and Kour, George and Lazar, Koren and Vetzler, Matan and Uziel, Guy and Anaby Tavor, Ateret},
  title     = {Effective Red-Teaming of Policy-Adherent Agents},
  booktitle = {Proceedings of the 2025 Conference on Empirical Methods in Natural Language Processing ({EMNLP})},
  pages     = {2250--2268},
  year      = {2025},
}

@inproceedings{levy2026stwebagentbench,
  author    = {Levy, Ido and Wiesel, Ben and Marreed, Sami and Oved, Alon and Yaeli, Avi and Mashkif, Nir and Shlomov, Segev},
  title     = {{ST-WebAgentBench}: A Benchmark for Evaluating Safety and Trustworthiness in Web Agents},
  booktitle = {International Conference on Learning Representations ({ICLR})},
  year      = {2026},
}

@inproceedings{ruan2024toolemu,
  author    = {Ruan, Yangjun and Dong, Honghua and Wang, Andrew and Pitis, Silviu and Zhou, Yongchao and Ba, Jimmy and Dubois, Yann and Maddison, Chris J. and Hashimoto, Tatsunori},
  title     = {Identifying the Risks of {LM} Agents with an {LM}-Emulated Sandbox},
  booktitle = {International Conference on Learning Representations ({ICLR})},
  year      = {2024},
}

@article{levi2025intellagent,
  author  = {Levi, Elad and Kadar, Ilan},
  title   = {{IntellAgent}: A Multi-Agent Framework for Evaluating Conversational {AI} Systems},
  journal = {arXiv preprint arXiv:2501.11067},
  year    = {2025},
}

@inproceedings{komoravolu2026ata,
  author    = {Komoravolu, Sameer and Mrini, Khalil},
  title     = {Agent-Testing Agent: A Meta-Agent for Automated Testing and Evaluation of Conversational {AI} Agents},
  booktitle = {Proceedings of the 19th Conference of the European Chapter of the Association for Computational Linguistics ({EACL})},
  pages     = {7199--7214},
  year      = {2026},
}

@inproceedings{sharma2026agentpex,
  author    = {Sharma, Reshabh K. and Barke, Shraddha and Zorn, Benjamin},
  title     = {Willful Disobedience: Automatically Detecting Failures in Agentic Traces},
  booktitle = {Proceedings of the {ACM} Conference on {AI} and Agentic Systems ({CAIS})},
  pages     = {584--596},
  year      = {2026},
}

@inproceedings{rabinovich2026nearmiss,
  author    = {Rabinovich, Ella and Boaz, David and Zwerdling, Naama and Anaby Tavor, Ateret},
  title     = {Near-Miss: Latent Policy Failure Detection in Agentic Workflows},
  booktitle = {Proceedings of the Fifth Workshop on Generation, Evaluation and Metrics ({GEM})},
  pages     = {296--308},
  year      = {2026},
}

@article{cuadron2025saber,
  author  = {Cuadron, Alejandro and Yu, Pengfei and Liu, Yang and Gupta, Arpit},
  title   = {{SABER}: Small Actions, Big Errors -- Safeguarding Mutating Steps in {LLM} Agents},
  journal = {arXiv preprint arXiv:2512.07850},
  year    = {2025},
}

\end{document}